\documentclass[12pt]{article}
\usepackage[english]{babel}

\RequirePackage{amsthm,amsmath,amsfonts,amssymb}
\RequirePackage[authoryear]{natbib}

\usepackage{adjustbox}
\usepackage{algorithm}

\usepackage{algpseudocode}
\usepackage{array}

\usepackage[english]{babel}
\usepackage{bigints}
\usepackage{booktabs}%,caption}

\usepackage{caption}
\usepackage{enumerate}
\usepackage{epstopdf}
\usepackage{graphicx,psfrag,epsf}

\usepackage{listings}
\usepackage{lscape}
\usepackage{multirow}
\usepackage{makecell}

\usepackage{rotating}
\usepackage[rightcaption]{sidecap}
\usepackage[flushleft]{threeparttable}
\usepackage{wrapfig}
\usepackage[table,xcdraw]{xcolor}

\usepackage[hidelinks,colorlinks=true,linkcolor=blue,urlcolor=blue, citecolor=blue,anchorcolor=blue]{hyperref}

\usepackage[top=1.9cm]{geometry}

\newtheorem{theorem}{Theorem}%[section]

\newtheorem{lemma}[theorem]{Lemma}

\newtheorem{innercustomgeneric}{\customgenericname}
\providecommand{\customgenericname}{}
\newcommand{\newcustomtheorem}[2]{%
  \newenvironment{#1}[1]
  {%
   \renewcommand\customgenericname{#2}%
   \renewcommand\theinnercustomgeneric{##1}%
   \innercustomgeneric
  }
  {\endinnercustomgeneric}
}

\usepackage{subfig}

\newcustomtheorem{customthm}{Theorem}
\newcustomtheorem{customlemma}{Lemma}

\providecommand{\keywords}[1]
{
  \small	
  \textbf{\textit{Keywords:}} #1
}

\title{Network method for voxel-pair-level brain connectivity analysis under spatial-contiguity constraints
}
\date{}
\author{\small Tong Lu$^{1}$, Yuan Zhang$^{2}$, Peter Kochunov$^{3}$, Elliot Hong$^{3}$, Shuo Chen$^{3,4}$\\
\footnotesize
$^{1}$Department of Mathematics, University of Maryland \\
\footnotesize
$^{2}$Department of Statistics, The Ohio State University\\
\footnotesize
$^{3}$Maryland Psychiatric Research Center, School of Medicine, University of Maryland, \\
\footnotesize
$^{4}$Division of Biostatistics and Bioinformatics, School of Medicine, University of Maryland, 
\footnotesize *\href{mailto:shuochen@som.umaryland.edu}{shuochen@som.umaryland.edu}
}

\vspace{-5mm}

\begin{document}
\maketitle
\begin{abstract}
Brain connectome analysis commonly compresses high-resolution brain scans (typically composed of millions of voxels) down to only hundreds of \textit{regions of interest} (ROIs) by averaging within-ROI signals. This huge dimension reduction improves computational speed and the morphological properties of anatomical structures; however, it also comes at the cost of substantial losses in spatial specificity and sensitivity, especially when the signals exhibit high within-ROI heterogeneity. Oftentimes, abnormally expressed \textit{functional connectivity} (FC) between a pair of ROIs caused by a brain disease is primarily driven by only small subsets of voxel pairs within the ROI pair. This article proposes a new network method for detection of voxel-pair-level neural dysconnectivity with spatial constraints. Specifically, focusing on an ROI pair, our model aims to extract dense sub-areas that contain aberrant voxel-pair connections while ensuring that the involved voxels are spatially contiguous.
%We achieve the goal by developing an entropy-driven cost function and a sub-community-detection algorithms that optimize the non-convex cost function via iterative procedures.
In addition, we develop sub-community-detection algorithms to realize the model, and the consistency of these algorithms is justified.
%The proposed method improves spatial specificity and biological interpretability for brain connectome analysis by providing precise disease-related connections on a voxel level.
Comprehensive simulation studies demonstrate our method's effectiveness for reducing the false-positive rate while increasing statistical power, detection replicability, and spatial specificity. We apply our approach to reveal: (i)~voxel-wise schizophrenia-altered FC patterns within the salience and temporal-thalamic network from 330 participants in a schizophrenia study; (ii)~disrupted voxel-wise FC patterns related to nicotine addiction between the basal ganglia, hippocampus, and insular gyrus from 3269 participants using UK Biobank data. The detected results align with previous medical findings but include improved localized information.

%schizophrenia heavily alters voxel-pair connections from the sub-areas of the dorsal insula and anterior cingulate cortices.
% (primarily comprising the cingulate cortex and bilateral insular cortices)
\end{abstract}
\keywords{Keywords: Brain connectome, spatial contiguity, voxel-pair-level connectivity, fMRI}

%Optimization; Bi-clique; Spatial constraints; Voxel-pair-level connectivity; Minimum description length; Ultra-high dimension; Brain connectome.

\section{Introduction}
Statistical network analysis and graph theory have been fundamental in the study of the intricate neural circuits in human brains (the ``human connectome'') \citep{bullmore2009complex,rubinov2010complex}. A large body of literature has revealed that the human connectome is a well-organized obscure network, and it exhibits graph properties of intelligent networks such as social networks and the Internet \citep{bahrami2019analysis,CAO201476}. Built on graph theory, brain network analysis depicts the brain connectome as a graph in which cortical regions are denoted as nodes and the connections between regions are edges. Under this framework, abundant statistical models have been developed to study the associations between complex neural connections and experimental/clinical conditions (e.g., \cite{simpson2013analyzing,fornito2016fundamentals}). These models can help to enhance our understanding of the underlying pathophysiological mechanisms of brain diseases (e.g., Alzheimer's disease and Parkinson's disease) and assist clinical predictions concerning disease diagnosis and treatment selection.
%\citep{zalesky2010network}.

\begin{figure}[htb]
\centerline{\includegraphics[width=1.05\textwidth]{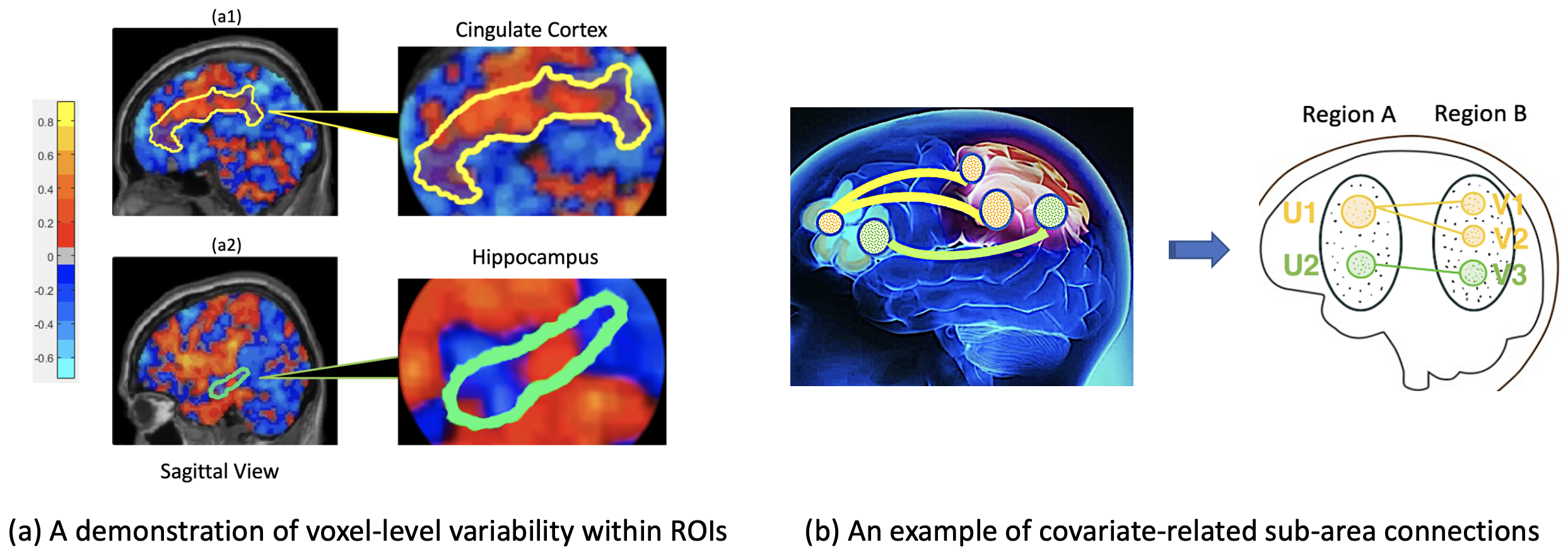}}
\caption{\footnotesize \textit{(a) shows the heterogeneity of functional connectivity(FC) among intra-ROI voxels from a seed-to-voxel analysis. Here, the insula was used as a seed ROI. The heatmaps in (a) characterize the FC between the seed ROI and voxels in the cingulate cortex (a1) or hippocampus (a2). Both cingulate and hippocampus are well-known ROIs, but their interior FC to insular varies substantially. (b) shows a simplified example of covariate-related FC between voxels in sub-area pairs $(U_1, V_1), (U_1, V_2), \textrm{and } (U_2, V_3)$ from a larger ROI pair (Region A, Region B).}}
\label{tab:ROI}
\end{figure}

In brain network studies, regions of interest (ROIs)
%(i.e., predefined clusters of anatomically close voxels)
are often considered as basic units of analysis, and these are equivalent to nodes/vertices in graph theory. The popularity of region-level brain network (RBN) analysis comes from its high anatomical consistency and computational tractability. When a whole-brain connectome is considered, RBN analysis dramatically reduces the search dimensions from trillions ($10^6\times10^6$) to thousands ($10^2\times 10^2$). However, RBN analysis relies on the assumption of signal homogeneity among intra-ROI voxels, which is often violated in reality. When significant intra-ROI heterogeneity is present, RBN analysis can lead to several analytical flaws:
\textit{Variability negligence}. Simply averaging the time series of voxels within an ROI can lead to voxel-level information variability loss (e.g., \autoref{tab:ROI}(a));
%which blurs detailed intra-regional information and deteriorate spatial resolution;
\textit{Spatial specificity loss}. A clinical covariate may alter the ROI-pair connections by disrupting only a small proportion of intra-ROI voxel pairs. In such cases, RBN analysis fails to precisely distinguish the localized alterations;
\textit{Power loss}. The averaging process mixes both significant and non-significant voxel-level connections, which often attenuates the effect size and statistical power.

Recently, many brain network studies have shifted focus from RBN analysis to voxel-level network analysis \citep{loewe2014fast,wu2013mapping}. Traditional multiple testing methods (e.g., the false-discovery rate (FDR) and the family-wise error rate (FWER) control) are not applicable to high-dimensional multivariate voxel pairs since they are unable to take into account anatomical restrictions and inherent systematical patterns of disease-associated voxels in ROIs. Some other existing methods may also have limitations, such as not utilizing rich voxel-level information to complement region-level connectivity characterization, or yielding relatively hard-to-interpret results for various reasons (e.g., under-represented neurobiological structures or biases in the seed-selection process).
%assist characterize/generalize region-level connection patterns.
% or they largely fail to unify the information regarding voxel-wise and region-wise connections
%they rarely combine information in voxel-wise and region-wise connections
Several advanced statistical methods have been proposed to address these limitations. For example, \citep{xia2017hypothesis} and \citep{xia2019matrix} provided localized statistical inference by accounting for the network properties. \citep{chen2016bayesian} proposed a Bayesian hierarchical model to identify the voxel-level connectivity patterns associated with clinical covariates and then used the voxel-wise functional connectivity (vFC) patterns to infer region-level connections. These novel approaches yield improved inference results and localized specificity. Nonetheless, they are not directly applicable to our input data of interest (i.e., an $m\times n$ ``bi-cluster'' rather than an $n\times n$ adjacency matrix), and they do not regulate involved voxels to be spatially contiguous. Unlike RBN analysis, spatial contiguity is crucial for vFC analysis because: (i)~it preserves anatomical homogeneity, and it hence preserves the interpretability of the vFC results \citep{thirion2006dealing}; (ii)~it better controls the FDR and FWER since phenotype-related vFC is often intrinsically linked with the topological structure of the brain connectome \citep{fan2012estimating}.

% These methods often yield improved inference results on individual edges by incorporating the edge-wise covariance structure. However, these results often face the issue of group-level inference, i.e., multiple testing on high-dimensional multivariate edges. If positive edges are randomly distributed in the whole brain, Xia and Li's methods as well as conventional multiple testing methods are well applicable because the location of edges can be randomly shuffled without impacting the inference results. However, this is often not the case in reality as abnormal edges are often distributed more systematically with complex topological structures. Moreover, these methods did not consider spatial adjacency constraints on the voxels; however, connections between spatially scattered voxels may often lead to less systematic and biological interpretations.

In this study, our goal was to identify altered vFC patterns between spatially contiguous sub-area pairs from a larger region pair. More specifically, given a region pair of interest, we sought to extract interior sub-area pairs that could maximally cover spatially adjacent covariate-related vFC with well-controlled FDR and FWER values (e.g., \autoref{tab:ROI}(b)). Our sub-area \textit{extraction} approach is fundamentally distinct from other commonly used brain \textit{parcellation} methods such as anatomy-based and data-driven approaches (e.g., gradient- or similarity-based mappings) \citep{wig2014approach,craddock2012whole}; these parcellation methods seek to segment an ROI into different sub-regions, and every single voxel is assigned to a corresponding sub-region.
%into several sub-regions, treating voxels as unlabeled and exchangeable nodes in a social network, without considering the spatial positions of voxels.
In contrast to parcellation methods in which every voxel is processed, our sub-area extraction approach only selects subsets of voxels that are covariate-related and are constrained in spatially contiguous spaces. All other non-selected voxels are considered to be covariate-indifferent. Sub-area extraction is more suited to our study because: (i)~it is likely that the covariate-related differences across clinical groups may gather in the vFC between a sub-area in Region~A and an intersection of multiple sub-areas grouped by the existing parcellation methods in Region~B; (ii)~it is often found that only a small proportion of voxels in regions~A and B are disrupted, and thus a comprehensive parcellation across the entire ROI is not necessary \citep{CAO201476}.

\begin{figure}[htbp]
\centerline{\includegraphics[width=1.1\textwidth]
{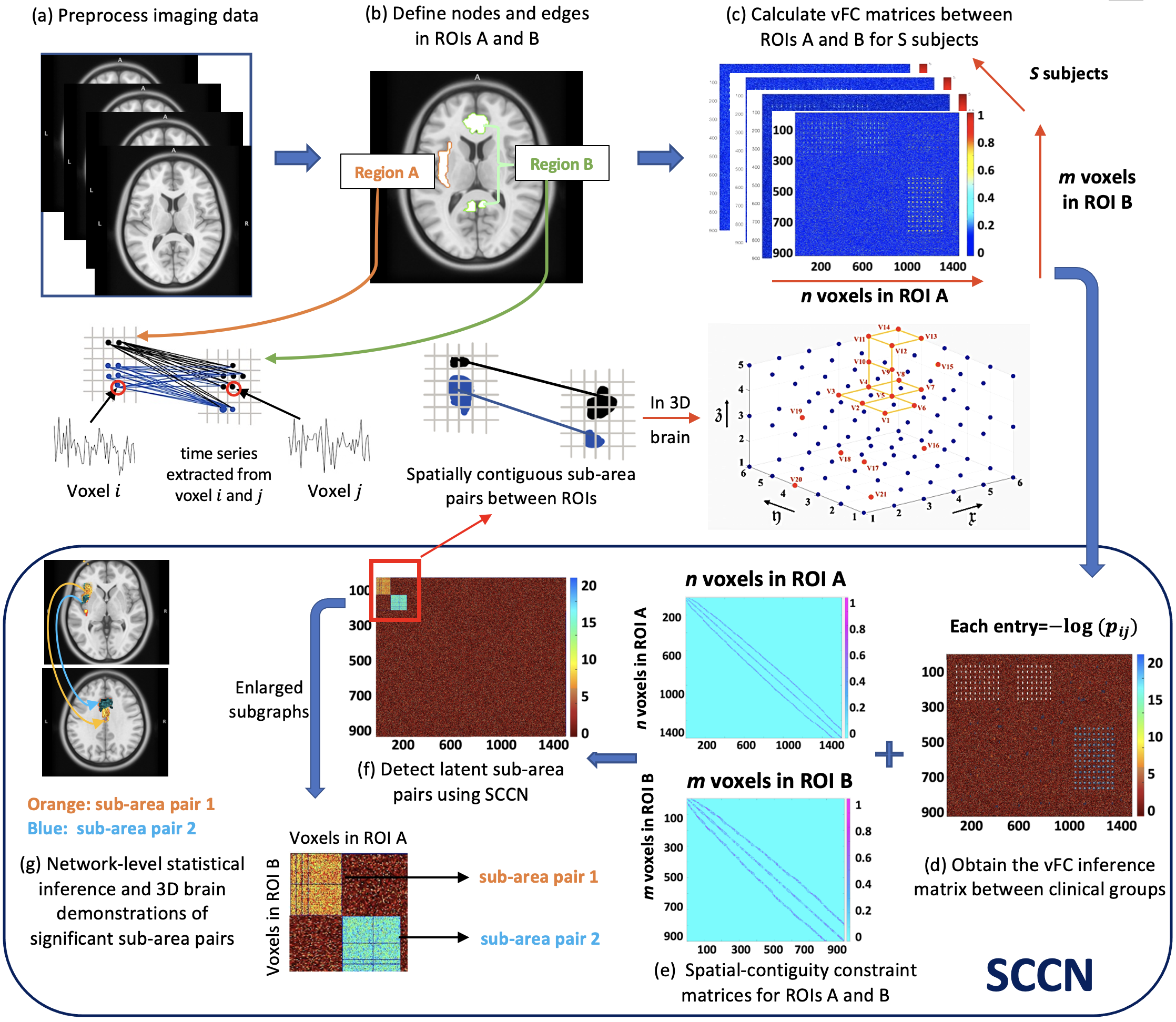}}

\caption{\footnotesize \textit{The vFC pattern extraction and inference pipeline. (a) Preprocess the fMRI data to remove unwanted artifacts and transform the data into a standard brain template. (b) Define voxels in ROIs as nodes and bonds between voxels as edges. Extract the time series of brain signals from each voxel. (c) Calculate the connectivity matrix between voxels from two regions, A and B, for each subject. (d) Calculate the connectivity inference matrix, where each element is a test statistic per edge between clinical groups. A hotter point in the heatmap suggests a larger between-group difference. (e) Calculate the spatial-contiguity constraint matrices for ROIs A and B (see detailed matrix construction in Section 2.1). 
In (e1), each dot represents a voxel in 3D coordinates, where red dots are voxels of interest. Voxels connected by yellow lines form a spatially contiguous area. (f) Detect the disease-related connections contained in sub-area pairs based on (d) and (e) jointly. (f) is obtained by re-ordering the nodes in (d), where the densely altered sub-networks are pushed to the top (i.e., (d) and (f) are isomorphic graphs). (g) Conduct the MDL-based network-level statistical inference. The sub-area pairs that pass the statistical tests are highlighted.
}}

\label{tab:pipeline}
\end{figure}

%To fill this methodological gap
To achieve the desired sub-area extraction and address the limitations discussed above, we propose a new statistical network framework to extract {\bf S}patially {\bf C}onstrained and {\bf C}onnected {\bf N}etworks, hereafter referred to as SCCN.
%which yields inference of both voxel- and region-level connectivity with improved power and interpretability.
SCCN is a two-step method (\autoref{tab:pipeline}) focusing between a pair of ROIs, say A and B, that are believed to contain aberrant functional connections caused by a brain disease. In \textbf{step~1}, SCCN extracts spatially coherent sub-area pairs that maximally contain disease-altered vFC between regions~A and B. In \textbf{step~2}, we formally test each extracted sub-area pair to determine whether it is significantly covariate-associated with multiple testing controls. If no sub-area pairs are found to be significant, we then consider the region-pair connectivity as covariate-unrelated. If significant results are seen, the association between the covariate of interest and the ROI-pair connections can be traced down to smaller but much more precise sub-areas consisting of extracted voxels. These vFC results may provide insights into understanding the latent neurophysiological mechanisms of diseases.

Herein, we show that SCCN provides a consistent estimate for the true community structure in the sense that the error of edge assignments is negligible in large region pairs. We also empirically evaluate the performance of SCCN using two real data examples and considerable simulation studies. In particular, the two real examples use functional magnetic resonance imaging (fMRI) data from a schizophrenia research study with 330 participants and a nicotine-addiction research study using UK Biobank\footnote{UK Biobank is a large-scale biomedical database and research resource containing in-depth genetic and health information from half a million UK participants.} data with 3269 participants.
For the schizophrenia data, we focus on the salience network and the temporal-thalamic network, both of which are well known to be susceptible to schizophrenic disorder. SCCN yields spatially coherent sub-area pairs in which most of the interior vFC is disrupted in schizophrenia, notably the (dorsal insula, anterior cingulate cortex) pair and the (anterior/medial temporal gyrus, medial thalamus) pair. For the UK Biobank data, we focus on the basal ganglia, hippocampus, and insular gyrus, which are believed to be vulnerable to the effects of nicotine. Here, SCCN yields spatially coherent sub-areas pairs containing addiction-associated vFC, notably the (medial inferior basal ganglia, medial inferior insula) pair. In contrast, traditional multiple testing correction methods (e.g., FDR and FWER control) and several existing RBN analyses have failed to reveal these findings with voxel-wise topological structures. Moreover, the results of our simulation studies also show that SCCN achieves satisfactory performance in increasing statistical power and spatial specificity while controlling the false-positive rate. In addition, SCCN is easily scalable to both small and large datasets.
%$\Leftrightarrow$

%%%%%%%%%%%%%%%%%%%%%%%%%%%%%%%%%%%%%%%%%%%%%%%%%%%%%%
\section{Methods}
\subsection{Background}%   \hfill \break
\subsubsection{Data structure}
Focusing on fMRI data, we want to investigate altered vFC patterns between two ROIs, A and B, consisting of $n$ and $m$ voxels, respectively. For a subject $s\in {[S]:=} \{1,\dots,S\}$, let $\mathbf{Z}^{A,s}_{n\times T}$ and $\mathbf{Z}^{B,s}_{m\times T}$ represent the matrices of voxel-level blood-oxygenation-level dependent (BOLD) signals at $T$ different time points for ROIs A and B. The outcome variables are the functional connectivity measures quantified by similarity matrices between the time series of voxels in A and in B. For example, $Y_{ij}^s$, the connectivity strength between voxel $i$ in A and voxel $j$ in B, can be computed by $Y_{ij}^s=f(Z_{i\cdot}^{A,s},Z_{j\cdot}^{B,s})$, where $Z_{i\cdot}^{A,s}$ and $Z_{j\cdot}^{B,s}$ are the BOLD time series for voxels $i$ and $j$, and $f$ is a similarity metric (e.g., Fisher's $z$-transformed Pearson correlation). Collecting all $Y_{ij}^s$ for each voxel pair $(i,j)\in [n]\times [m]$ gives an inter-region connectivity matrix $\mathbf{Y}_{n\times m}^s$. Additionally, a covariate vector $\mathbf{X}^s_{1\times p}$ is observed for each subject $s$, and this contains demographic and clinical information.

Our goal is to identify clinical/behavioral-related functional connectivity (FC) patterns at the voxel level. This is because voxel-level findings can reveal altered FC with improved statistical power and enhanced spatial specificity and resolution. To achieve this, multivariate statistical inference is required for the $n \times m$ vFC outcomes (usually in high dimension, e.g., millions) with spatial constraints. We first test the associations between each outcome $Y_{ij}^s$ and a regressor of primary interest $x_1^s \in \mathbf{X}^s$ (clinical status in our application, e.g., patient or control):  
\begin{align*}
\mathop{\mathbb{E}}(Y_{ij}^s|\mathbf{X}^s) =\alpha_0+x_1^s\beta_{ij}+\mathbf{X}_{1\times (p-1)}^s \mathbf{\boldsymbol \alpha},
\end{align*}
where $\beta_{ij}$ is the coefficient of $x_1^s$ and $\mathbf{\boldsymbol \alpha}$ is a coefficient vector for the remaining covariates $\mathbf{X}_{1\times(p-1)}^s$ (e.g., age, ethnicity, etc). We denote $\mathbf{\boldsymbol \beta}:=\{ \beta_{ij}\}_{i\in [n], j\in [m]}$ and aim to systematically extract vFC whose $\boldsymbol{\beta}\neq 0$ with high accuracy. We further summarize the significance levels of $\mathbf{\boldsymbol \beta}$ by a \textit{connectivity inference matrix} $\mathbf{W}_{n \times m}$. Each entry of $\mathbf{W}_{n \times m}$ is computed by $\mathbf{W}_{ij} = -\log p_{ij}$, where $p_{ij}$ is the $p$-value for $\beta_{ij}$. In neuroimaging statistics, the selection of $\boldsymbol{\beta}\neq 0$ is not only determined by the level of statistical significance but also by spatial constraints. In addition to these two factors, $\mathbf{\boldsymbol \beta}$ is also intrinsically linked with an underlying $n \times m$ bipartite graph between ROIs A and B. Therefore, we will require both graphic and spatial information to assist in identifying vFC whose $\boldsymbol{\beta}\neq 0$. We present the detailed graphic and spatial constructions as follows.

% \revision{
% Our goal is to investigate network-level associations between vFC captured in $\mathbf{Y}_{n\times m}^s$ and covariates $\mathbf{X}^s_{1\times p}$ so that the inherent covariate-related vFC patterns of topological structure between ROIs A and B can be revealed. To do this, we first model the associations between $\mathbf{Y}$ and $\mathbf{X}$ as
% \begin{align*}
% \mathop{\mathbb{E}}(Y_{ij}^s|\mathbf{X}^s) =\alpha_0+x_1^s\boldsymbol{\beta}_{ij}+\mathbf{X}_{1\times (p-1)}^s \mathbf{\boldsymbol \alpha},
% \end{align*}
% where $\boldsymbol{\beta}_{ij}$ is the coefficient for the covariate of primary interest such as disease status (e.g., patient or control), and $\mathbf{\boldsymbol \alpha}$ is the coefficient vector corresponding to the rest $(p-1)$ covariates (e.g., age, race, etc). $\mathbf{\boldsymbol \beta}:=\{\boldsymbol{\beta}_{ij}\}_{i\in [n], j\in [m]}$ is usually of ultra-high dimension ($n\times m$ can be millions), and studying voxel pairs containing $\boldsymbol{\beta}\neq 0$ provides essential knowledge in identifying covariate-related vFC patterns. To further reveal these patterns with systematic network structures, we need to jointly study $\mathbf{\boldsymbol \beta}$ with additional components including inference information of $\mathbf{\boldsymbol \beta}$, the underlying graph structure and spatial constrains. All these components are introduced as follows.
% }

%\vspace{2mm}

\subsubsection{Graph representation}
To decipher the complex voxel-pair connectome, we consider a bipartite graph structure $G=\{U, V\}$ underlying the inference matrix $\mathbf{W}_{n\times m}$. The node sets $U$ and $V$ represent voxels in ROIs A and B, respectively, where $|U|=n$ and $|V|=m$. We assume that, after spatial normalization and registration of the fMRI data, all subjects share common node sets, namely, $(U^s,V^s)\equiv (U,V), \forall s\in [S]$.
%The edge set $E$ represents vFC between $U$ and $V$, where $|E|=n\times m$.
%For each edge $e_{ij}\in E$, we have an indicator variable $\delta_{ij}=1$ if ${\beta}_{ij} \neq 0$ and $\delta_{ij}=0$ otherwise. We estimate $\hat{\delta}_{ij}$ based on $\mathbf{W}_{n\times m}$ by our proposed network approach described in the following section.

%\vspace{2mm}

\subsubsection{Spatial contiguity}
Each node in our dataset corresponds to a voxel at a certain spatial position in 3D brain imaging (e.g., \autoref{tab:pipeline}(e1)). When we map each detected subgroup of voxels back to the 3D brain space, we desire these voxels to emerge as a spatially adjacent cluster (i.e., connected components). Such anticipation, translated into formal language, is referred to as \emph{spatial contiguity}. Specifically, we define an ``infrastructure graph'' ${\cal S}_A$ between all nodes within ROI~A to accommodate spatial contiguity. Each entry $S_{ii'}$ in ${\cal S}_A$ is a spatial-adjacency indicator variable between voxels ${i}$ and ${i'}$ in ROI~A, where $S_{ii'}=1$ if $d_{ii'} \leq \varepsilon$, and $S_{ii'}=0$ otherwise ($d_{ii'}$ is the Euclidean distance between voxels $i$ and $i'$). For example, in a 3D grid space, when $\epsilon$ is set to be $\sqrt{3}$, a centroid voxel $i$ in a cube will have 26 surrounding voxels $i'$ such that $S_{ii'}=1$. We define and interpret ${\cal S}_B$ for nodes within ROI~B similarly. ${\cal S}_A$ and ${\cal S}_B$ will be used to prescribe the spatial-contiguity constraints when implementing SCCN. We provide more rigorous mathematical definitions of spatial contiguity, ${\cal S}_A$, and ${\cal S}_B$ in Appendix~A.1.

%\vspace{3mm}
We propose the SCCN model to systematically select vFC of ${\beta}_{ij} \neq 0$ by jointly considering the information of voxel-pair-level statistical significance, underlying graph structures, and spatial constraints. We integrate these into a weighted graph ${\cal G} = \{W,{\cal S}_A,{\cal S}_B\}$ as the input of our method.

% \tong{\autoref{tab:spa_con} shows an example of a 3D structure of brain voxels from an ROI, where neuroimaging data is often stored in grid structures as shown.
% Voxels in yellow indicate disease-related voxels. By choosing $d_{ii'}=\sqrt{3}$, V1 and V2 are considered as spatially contiguous, and V1-V14 form a spatially contiguous sub-network, while V15-V21 is not spatially contiguous to this sub-network or any voxels in this sub-network.}

% \begin{figure}[!ht]
% \centerline{\includegraphics[scale=.45]{pics/spa_con copy.png}}
% \caption{\footnotesize An example of within-region voxel-wise connection structure.}
% \label{tab:spa_con}
% \end{figure}

% \vspace{5mm}
% In summary, the input data for our method are ${\cal G} = \{W,{\cal S}_A,{\cal S}_B\}$. Usually, no prior knowledge is available for how voxel-level dysconnectivity patterns precisely exhibit between a ROI pair. We are motivated to develop a valid network model to extract the latent bipartite subgraph sets within $(U,V)$ and perform network-level statistical inference on the extracted subgraph sets while well controlling FDR and FWER.

\subsection{Detecting densely altered sub-area pairs from an ROI pair}
\subsubsection{Spatial-contiguity-constrained objective function}
The node set $U$ corresponding to voxels in ROI~A can reportedly be partitioned into mutually non-overlapping sub-areas $\{U_c\}$, denoted by $U = \bigoplus_{c=1}^C U_c$ \citep{eickhoff2015connectivity}. Similarly, we have $V = \bigoplus_{d=1}^D V_d$ for ROI~B. In this paper, we aim to extract sub-area pairs $\{(U_c,V_d)\}$ that dominantly contain disease-related voxel pairs, and we call these ``densely altered'' sub-area pairs.
%They can effectively reflect vFC patterns associated with clinical conditions of interest.
Formally, a sub-area pair $(U_c,V_d)$ is considered densely altered if $  \sum_{(i,j)\in (U_c, V_d)  } \frac{I(\beta_{ij}\neq 0)}{|U_c|~|V_d|}
    \gg 
    \sum_{(i,j)\in (U'_c, V'_d) } \frac{I(\beta_{ij}\neq 0)}{|U'_c||V'_d|}
$, where $U'_c$ and $V'_d$ are the complements of node sets $U_c$ and $V_d$.
%This aligns with the intuition of Lemma~1 in \citept{wu2021multivariate} and also \citept{wu2021extracting}.
We are therefore inspired to devise a regularized objective function to generate a checkerboard-like network structure underlying the connectivity inference matrix $\mathbf{W}$. This network structure reshuffles $\mathbf{W}$ and reveals densely altered $\{(U_c,V_d)\}$ pairs from $(U,V)$. In addition, we impose spatial contiguity on $U_c$ and $V_d$ to improve biological interpretability and prohibit isolated false positive edges. Finally, the objective function is formulated as follows:
%This community structure will be a cornerstone that enables us to further identify the densely altered sub-area pairs via cluster-wise statistical inference proposed in Section \ref{test}. The objective function based on $\mathbf{W}$ is formulated as follows:
\begin{align}
     \label{eq:obj}
     \underset{
        \substack{
            C, ~D,
            ~U = \bigoplus_{c=1}^{ C} U_c,
            V = \bigoplus_{d=1}^{D} V_d
            \\
             (U_c, V_d \text{ subject to spatial contiguity})
        }
     }
     {\mathrm{argmax}}\,
     \bigintsss
     \sum_{c=1}^C\sum_{d=1}^D
     \Bigg\{
        \log&\frac{\sum_{i\in U_c, j\in V_d}W_{ij}\cdot I(W_{ij} > r)} {|U_c||V_d|}\\
& + \lambda \log(|U_c||V_d| )
     \Bigg\}~g(r) dr, \nonumber
\end{align}
where $\lambda \in [0,1]$ is a tuning parameter, $r$ is a threshold below which there is no disease-related effect on $W_{ij}$, and $g(r)$ is the distribution function for $r$. Both $g(r)$ and $\lambda$ can be chosen by prior knowledge or by a data-driven method proposed in Section~\ref{datadriven}.

The tuning parameter $\lambda$ falls in the range $[0,1]$: when $\lambda=0$, maximizing \eqref{eq:obj} is equivalent to maximizing $f_1= \frac{\sum_{i\in U_c, j\in V_d}W_{ij}\cdot I(W_{ij} > r) } {|U_c||V_d|}$, which is a popular definition for connection density; when $\lambda=1$, maximizing \eqref{eq:obj} is simply maximizing $f_2=\sum_{i\in U_c, j\in V_d}W_{ij}\cdot I(W_{ij} > r)$, which quantifies the magnitude of significant voxel pairs contained by the sub-area pair $(U_c, V_d)$. Direct optimization of the connection density $f_1$ tends to detect a dense subgraph with a minuscule size, while the optimization of $f_2$ can trigger an oversized subgraph. Theorem \ref{th1} shows that function~\eqref{eq:obj} provides a consistent estimate for the targeted topological structure (collections of edge-induced sub-area pairs) in the sense that the error of edge assignments is negligible in large region pairs. Extensive simulation studies also show that function~\eqref{eq:obj} performs well in balancing the size and density when detecting subgraphs.

% The main distinction between our proposal \eqref{eq:obj} and similar counterparts in previous literature \citep{chen2019graph, wu2021multivariate} lies in that function \eqref{eq:obj} honors spatial contiguity, which significantly improves biological interpretability and further suppresses isolated false-positive edges in the estimation result. Also, Function \eqref{eq:obj} enables us to extract signal edges between regions, which overcomes the limitation of within-region extraction only. Moreover, the input inference matrix $\mathbf{W}$ accounts for not only the voxel-based connectome information of regions A and B but also each patient's profiling covariates.

\subsubsection{\texorpdfstring{Optimization of objective function \eqref{eq:obj} for given $g(r)$ and $\lambda$}{Optimization of objective function}}
\label{datadriven}
%Our narration of this section is outlined as follows:
In this section, we focus on optimizing function~\eqref{eq:obj} for a given configuration of $g(r)$ and $\lambda$, which are the density function for the threshold $r$ and the tuning parameter in \eqref{eq:obj}. We will then discuss how to determine $g(r)$ and $\lambda$ in the next section. Unfortunately, even with a given $g(r)$ and $\lambda$, direct optimization of \eqref{eq:obj} is still an NP-hard problem. Therefore, traditional optimization methods, such as gradient descent, cannot be used due to the non-convexity of the problem. Here, we present an alternative strategy for approaching \eqref{eq:obj}. The essential idea is that we integrate $\mathbf{W}$ with the spatial-contiguity constraints and then estimate the targeted community structure using modified spectral clustering algorithms via iterative procedures. As presented earlier, the targeted network structure is $\{U_c, V_d\}$ partitioned from $(U, V)$ (i.e., the collection of edge-induced sub-area pairs, or in other words, the voxel memberships of $U_c$ and $V_d$), where $U=\bigoplus_{c=1}^C U_c$ and $V=\bigoplus_{d=1}^D V_d$.
%\citep{gao2017achieving}

%Spectral clustering on $\mathbf{W}$ is essentially linked to performing eigendecompositions on the Laplacian matrices of  $\mathbf{W}\mathbf{W}^\top$ and $\mathbf{W}^\top\mathbf{W}$.
According to the spectral clustering algorithm, applying singular value decomposition to the Laplacian matrix of $\mathbf{W} = \mathbb{U} \mathbf {\Sigma} \mathbb{V}^\top$ and then clustering $\mathbb{U}$ and $\mathbb{V}$ will give partitions of regions~A and B, respectively. Now, since $\mathbb{V}$ is the eigenvectors of $\mathbf{W}^\top \mathbf{W}$, spectral clustering on the Laplacian matrix of $\mathbf{W}^\top \mathbf{W}$ will simply give the partitions of Region~B. Similarly, spectral clustering on the Laplacian matrix of $\mathbf{W} \mathbf{W}^\top$ will provide the partitions of Region~A. Therefore, our community-detection algorithm can be conducted based on $\mathbf{W}\mathbf{W}^\top$ and $\mathbf{W}^\top\mathbf{W}$.
Next, to incorporate the spatial-contiguity constraints into the optimization, we make use of the two within-region ``infrastructure graphs'' ${\cal S}_A$ and ${\cal S}_B$ introduced earlier. Specifically, we define
%\citep{cano2007possibilistic}
\begin{align}
\label{eq:SaSb}
    \mathbf{W_A}
    &=~
    \mathbf{W}\mathbf{W}^\top \odot {\cal S}_A
    \quad
    \textrm{and}
    \quad
    \mathbf{W_B}
    =~
    \mathbf{W}^\top \mathbf{W} \odot {\cal S}_B,
\end{align}
where $\odot$ is an element-wise product. As pointed out by \cite{ilprints587} and \cite{craddock2012whole}, ${\cal S}_A$ and ${\cal S}_B$ force the similarity between all pairs of non-adjacent voxels to zero, which breaks edges between isolated voxels in the graph. Based on this, the $n$ by $n$ matrix $\mathbf{W}_A(ii')$ (where $i$ and $i'$ are two voxels in A) is greater if the voxels in A are spatially adjacent and have a similar profile linking to voxels in Region~B.
%In empirical studies, our method always extracts spatially contiguous sub-area pairs $\{(U_c, V_d)\}$ that contain far higher densities of covariate-related edges.
The spatial-contiguity constraints enable our method to produce results that better honor the neurobiological background regarding the coherence of neighboring neuron populations \citep{thirion2006dealing}.

We can now fit a stochastic block model to $\mathbf{W}_A$ (and another to $\mathbf{W}_B$) using the spectral clustering algorithm and then grid search for the optimizer of function~\eqref{eq:obj}.
%This produces $\{\hat U_c\}_{c=1}^C$ and $\{\hat V_d\}_{d=1}^D$, which we use to substitute in Function \eqref{eq:obj}.
We further examine whether the estimated $\hat U_c$ and $\hat V_d$ values satisfy the spatial-contiguity constraints, while empirically we find that the constraints are typically satisfied. There is thus no need to perform any further modification step for the constraints. We formally present our clustering procedure in Algorithm~\ref{alg:alg1}.

\begin{algorithm}
    \caption{Optimization of objective function~\eqref{eq:obj} with given $\lambda$}.
    \label{alg:alg1}
    \begin{algorithmic}[1]
    \Procedure{Algorithm }{Input: $\lambda$ and $\mathcal{G}=\{\mathbf{W},\cal{S}_A,\cal{S}_B\}$}
        \State \textbf{function} \textit{SCCN.partition} ($\lambda$, $\mathcal{G}$ )
    	   \For {$C=1,2,\ldots, |U|$}
    	        \State Ratio-cut spectral clustering $\mathbf{W}_A$ into $C$ networks:
    			$U = \bigoplus_{c=1}^C U_c$
    			\For {$D=1,2,\ldots,|V|$}
    			\State Ratio-cut spectral clustering $\mathbf{W}_B$ into $D$ networks:
               $V = \bigoplus_{d=1}^D V_d$
                \State Substitute network sets $U$ and $V$ into objective function~\eqref{eq:obj}, and obtain the output values %SCCN-value ($K_A$,$K_B$)
    			\EndFor
    	    \EndFor
        \State \textbf{return} $C, D$, $U = \bigoplus_{c=1}^C U_c$ and $V = \bigoplus_{d=1}^D V_d$ that yield the maximum output value
        \State \textbf{end function}
    \EndProcedure
    \end{algorithmic}
\end{algorithm}

\textbf{Consistency for subgraph detection.}
In Lemma\ref{lemma1}, we first establish that, given true sub-areas numbers $C^*$ and $D^*$, the solution to optimize the objective function~\eqref{eq:obj} provides a consistent estimate for the topological structure of the target community $(\{U_c, V_d\})$ (the collection of edge-induced sub-area pairs) in the sense that false-positive edge assignments are negligible in very large bipartite graphs $G=(U,V)$, $|U|\to \infty$, and $|V|\to \infty$.
%Furthermore, we establish in Theorem \ref{th1} that the community structure based on the sub-area numbers $\hat C, \hat D$ selected by grid search is also consistent.
Then, we establish the convergence of Algorithm~1 to optimize function~\eqref{eq:obj} based on Theorems \ref{th1} and \ref{th2}. In Theorem \ref{th1}, we prove that our algorithm can provide a consistent estimate of the number of sub-areas, C and D. In Theorem \ref{th2}, we prove that the implementation of Algorithm~1 converges to the optimal solution of the objective function~\eqref{eq:obj}.

%which verifies the resilience of Algorithm \ref{alg:alg1}.  }

To present the theoretical results, we consider the following settings. Let $\{e_{ij}^1\}$ and $\{e_{ij}^0\}$ be the sets of positive (e.g., disease-related) and negative edges, respectively. For an adjacency matrix $\mathbf{W}$, we assume that $w_{ij}|e_{ij}^1 \overset{\mathrm{iid}}{\sim} f_1$ and $w_{ij}|e_{ij}^0 \overset{\mathrm{iid}}{\sim} f_0$, where $f_1$ and $f_0$ are two probability density functions with means and variances $(\mu_1, \sigma_1^2)$ and $(\mu_0, \sigma_0^2)$, respectively. In addition, let $\mathcal{M}^*$ be the true membership of edges (the community index of edges falling in sub-area pair $(U_c, V_d)$). Furthermore, let $\hat{\mathcal{M}}_{(\hat C,\hat D)}$ be the membership estimated by function~\eqref{eq:obj} with $\hat C$ sub-areas in Region~A and $\hat D$ sub-areas in Region~B.

\begin{lemma}
\label{lemma1}
%\revision{
(\textbf{Consistency with known sub-area numbers $C^*$ and $D^*$}). 
Assume that $\mathbb{E}(\mathbf{WW^T})$ is of rank $C^*$ with smallest absolute nonzero eigenvalue of at least $\Lambda_A$, and  $\mathbb{E}(\mathbf{W^TW})$ is of rank $D^*$ with smallest absolute nonzero eigenvalue of at least $\Lambda_B$ . Assume further that $\mathrm{max}(\mu_0,\mu_1, \sigma_0^2,\sigma_1^2)\leq d$ for some $d\leq \mathrm{max}(\mathrm{log}n/n,\mathrm{log}m/m))$. Then, if there exists $(2+\varepsilon_A)\frac{ndCD}{\Lambda_A^2} < \tau_A$ and $(2+\varepsilon_B)\frac{mdCD}{\Lambda_B^2} <\tau_B$ for some $\tau_A, \tau_B, \varepsilon_A, \varepsilon_B>0$, the output $\hat{\mathcal{M}}_{(\hat C,\hat D)}$ that maximizes function \eqref{eq:obj} is consistent to the true membership $\mathcal{M}^*_{(C^*,~D^*)}$ underlying the latent community structure up to a permutation. 
%}

%\revision{
Equivalently, let $\hat{S}_c$, $\hat{S}_d$ be the estimated node sets for the subgraphs $G_c, G_d$ (induced by $U_c$ and $V_d$), respectively. Then $\hat{S}_c \cap U_c$ represents the nodes in $G_c$ whose assignments can be guaranteed. $\hat{S}_d \cap V_d~$ follows the same definition. With probability at least $1-\mathrm{max}(n,m)^{-1}$, up to a permutation, we have 
\begin{align}
    \sum_{c=1}^C \sum_{d=1}^D \bigg[1-\frac{ \big|~(\hat{S}_c \cap U_c)\bigotimes (\hat{S}_d \cap V_d) ~\big| }{|U_c|~|V_d|}\bigg]
    \leq \tau_A^{-1}(2+\varepsilon_A)\frac{ndCD}{\Lambda_A^2} +
    \tau_B^{-1}(2+\varepsilon_B)\frac{mdCD}{\Lambda_B^2},
    \notag
\end{align}
where $\bigotimes$ denotes the edge set that connects two node sets on its left and right side.
%}

\end{lemma}

\begin{customthm}{1}\label{th1}
%\revision{
(\textbf{Consistency for grid-searched C, D}). Let the sizes of subgraph pairs $|U_c|\times|V_d| (\forall c=[C^*], d=[D^*])$ be generated from a multinomial distribution with probabilities $\mathbf{\pi}=(\pi_1, \dots,\pi_{C^*\times D^*})$. Assume $~\exists \delta >0$, such that
\begin{align}
\notag
    \mu_1>\mu_0 \frac{1+\delta}{1-\delta}\bigg(1+\sqrt{1+\frac{\pi^2_{\mathrm{min}}}{\pi^2_{1}+\dots+\pi^2_{C^*\times D^*}}}\bigg),
\end{align}
then under conditions in Lemma \ref{lemma1} and tuning parameter $\lambda=0.5$, the number of mis-assigned edges $N_{\mathrm{edge}}$ satisfy 
\begin{align}
\notag
    N_{\mathrm{edge}}=o_p(n_{\mathrm{min}}*m_{\mathrm{min}}) \mathrm{as} ~|U|,|V|\rightarrow \infty,
\end{align}
where $n_{\mathrm{min}}, m_{\mathrm{min}}$ are the sizes of the smallest possible subgraphs in $U$ and $V$, respectively.
%}
\end{customthm}

\begin{customthm}{2}\label{th2}
%\revision{
(\textbf{Convergence of Algorithm \ref{alg:alg1}}). Let $\Tilde{U}=\bigoplus_{c=1}^{\Tilde{C}} U_c, \Tilde{V}=\bigoplus_{d=1}^{\Tilde{D}} V_d$ be
the partitions yielded by ratio-cut spectral clustering on $\mathbf{W}_A$ and $\mathbf{W}_B$ that maximizes function (1) with cluster numbers $\Tilde{C},\Tilde{D}$. Then $\Tilde{U},\Tilde{V}$ converge almost surely to the true community structure where false-positive edge assignments to each sub-bicluster are negligible.
%}
\end{customthm}

\textit{Proof}. Proofs of Lemma \ref{lemma1} and theorems \ref{th1} and \ref{th2} are provided in Appendix~B.

\vspace{1mm}

In summary, the above results provide theoretical evidence that the solution of the proposed objective function~\eqref{eq:obj} and Algorithm~\ref{alg:alg1} converge to the target community structure $(\{U_c, V_d\})$. Moreover, extensive simulation analyses in multiple settings with a wide range of different sample sizes demonstrate that SCCN can accurately reveal the true community structure with low false-positive and false-negative rates.

\subsubsection{\texorpdfstring{Determining $g(r)$ and $\lambda$}{Determining \textit{g}(\textit{r}) and lambda}}
{\bf \emph{Determining $g(r)$.}} Following \citep{efron2012large}, we can choose $g(\cdot)$ to be a discrete distribution on thresholds $\{r_1,\ldots,r_p\}$. A simple example would be as follows. Suppose that the voxel-pair-level FDRs yielded by pre-selected thresholds $r_1$, $r_2$, and $r_3$ are $0.20$, $0.10$, and $0.05$, respectively. We can then assign a higher probability mass to $r_p$ that yields a lower FDR, for example, $g(r_1)=0.1$, $g(r_2)=0.3$, and $g(r_3)=0.6$. In addition, $r_1$, $r_2$, and $r_3$ can be chosen from commonly used thresholds in MRI studies, such as $-\log(0.005)$ and $-\log(0.001)$ ($W_{ij}$ is $-\log p_{ij}$ after screening, and $r$ is a threshold for $W_{ij}$).

\vspace{1.5mm}

{\bf \emph{Selecting $\lambda$.}} As aforementioned, the tuning parameter $\lambda$ adjusts the balance between the subgraph size and the connection density; it thus plays a critical role in our method. A large $\lambda$ encourages large $|U_c|$ and $|V_d|$, whereas a small $\lambda$ is stricter on the connection densities of $(U_{c},V_d)$ pairs. Essentially, the selection of $\lambda$ is related to the network structure of ${\beta}_{ij}$. In practice, we have observed from many datasets that the coefficient ${\beta}_{ij} \neq 0$ usually exhibits a block model. To reflect this, we assume the following hierarchical model. Suppose there exists a non-random, latent hyperparameter $\boldsymbol{\beta} \in \mathbb{R}^{n\times m}$ with all nonzero elements. We can generate a bipartite similarity matrix $\eta\in\{0,1\}^{n\times m}$ from a bipartite stochastic block model with blocks $\{(U_c, V_d)\}$ and the corresponding connection probabilities $\{\pi_{ij}\}$, such that $\eta_{ij}\sim \mathrm{Bernoulli} (\pi_{ij})$ are independent of each other, where
    %The working $\boldsymbol{\beta}_1=\{\beta_{(i,j)_1}\}$ coefficient in our model is then generated by $\beta_{(i,j)_1} = \delta_{i,j}\tilde\beta_{(i,j)_1}$.
\begin{equation*}
\pi_{ij}  = \left\{
        \begin{array}{ll}
            \pi_{cd}(\lambda) & \quad i\in U_c, j\in V_d ,\\
            \pi_0(\lambda) & \quad \text{otherwise.}
        \end{array}
    \right.
\end{equation*}
We select the $\lambda$ value that maximizes the likelihood for this block model. In practice, the $\eta_{ij}$ values are not directly observable, and we replace them by $\eta_{ij}(r_0):= I(w_{ij}>r_0)$. The log-likelihood function for $\lambda$ is:
\begin{align}
%\label{eq:lambda1}
\notag
%\small % Please note that this command is invalid in math mode
    l_\lambda( \pi_{cd},  &\forall c\in[C], d\in[D] |\eta_{ij}(r_0))
    = \sum_{c,d} \sum_{(i,j) \in U_c \times V_d }  \eta_{ij}(r_0)\log{\pi_{cd}} 
&+ (1-\eta_{ij}(r_0)) \log \left(1-\pi_{cd}\right). 
\end{align}

To eliminate the arbitrariness in choosing the threshold $r_0$, we integrate the likelihood function with respect to $r_0$ over a prior distribution $g_0(r_0)$ determined by the method above.
%Similar to modeling the density function $g(r)$ for the threshold $r$ in function~\eqref{eq:obj}, we can set $g_0(\cdot)$ for $r_0$ to be a discrete distribution on the support $\{(r_0)_1,\ldots,(r_0)_q\}$.
This yields the following criterion:
\begin{align}
\label{eq:gamma2}
\notag
%\small % As noted above, this command is invalid in math mode
        \lambda_{\rm optimal}
        =
        {\rm argmax}_\lambda
        \Big\{
            &\int
            \max_{U_c,V_d,\pi_{cd}}
            %\notag\\
            &l_\lambda^r \left(\pi_{cd}(\lambda), \forall c=[C], d=[D] ~|~ \eta_{ij}(r_0) \right) ~g_0(r_0) d r_0
        \Big\}.
\end{align}

We formally present the procedure to select the tuning parameter $\lambda$ in Algorithm~\ref{alg:alg2}. The overall complexity of the algorithm is $O(Knm)$, where $K$ is a sufficient searching range for $\lambda$, $n=|U|$, and $m=|V|$.
%The convergence properties of Algorithm~\ref{alg:alg2} are discussed in \textbf{Appendix~B.3}.
Since the inference results between clinical groups across $S$ subjects are captured in $\mathbf{W}$, the algorithm complexity no longer involves sample size $S$, indicating the scalability of SCCN for large datasets. In addition, clustering algorithms typically involve computing the first $K$ eigenvectors of a potentially high-throughput similarity matrix. Our input similarity matrix $\mathbf{W}$ is sparse after applying screening and the spatial-contiguity constraints (usually only 0.2\%--5.0\% of edges are non-zero entries after processing), which notably reduces computational expense. It is, however, worth noting that since our algorithm is based on a single region pair, the computational burden may become heavy when investigating multiple different pairs, especially when a whole-brain analysis is needed.

\begin{algorithm}[ht!]
\caption{Grid search for $\lambda$.}
\label{alg:alg2}
\begin{algorithmic}[1]
\Procedure{Algorithm }{}
    \For {$0\leq \lambda \leq 1$}
        \State \textbf{return} $U = \bigoplus_{c=1}^C U_c$ and $V = \bigoplus_{d=1}^D V_d$ by Algorithm~\ref{alg:alg1}
        \For {$r_0=(r_0)_1$ to $(r_0)_q$}
           \State Compute the log-likelihood: $l_\lambda(\hat{\pi}_{c\times d}^{\mathrm{MLE}}, ~ \forall c=[C], d=[D] ~|~ \eta_{ij}(r_0))$
        \EndFor
        \State Integrate the log-likelihood w.r.t. $r_0$:
        \State $l_\lambda
        = \sum_{i=1}^p L_\lambda \left(\hat{\pi}_{c\times d}^{\mathrm{MLE}}, \forall c=[C], d=[D] |~ \eta{ij}(r)\right) ~g((r_0)_i)$
    \EndFor
%    \vspace{3mm}
    \State \textbf{return} $\hat{\lambda}$ that yields maximized $l_\lambda$

\EndProcedure
\end{algorithmic}
\end{algorithm}

\subsection{\texorpdfstring{Statistical inference of $\{(U_c,V_d)\}$ pairs}{Statistical inference of \{(Uc,Vd)\} pairs}} \label{test}
Recall that our ultimate goal is to extract a few most-densely connected subgraph pairs from $\{(U_c,V_d)\}$ based on the block partition $\{U_c, V_d: c\in[C], d\in[D]\}$ that we have already obtained at this point. A natural idea is to inspect each $(U_c,V_d)$ pair and perform a statistical test on them with the alternative hypothesis that the subgraph $U_c\otimes V_d$ is unusually dense. Here, we devise a cluster-wise permutation test \citep{nichols2002nonparametric} with FWER control. The hypotheses are:
\begin{align*}
    H_0&: \textrm{Subgraph $U_c\otimes V_d$ is not unusually dense},\\
%    \quad
    H_a&: \textrm{Subgraph $U_c\otimes V_d$ is unusually dense}.
\end{align*}
More specifically, under $H_0$, the connection density of $U_c\otimes V_d$ should be close to the density of sub-area pairs obtained by randomly shuffling edges in the bipartite graph. Similar to \citep{grunwald2007minimum}, we devise the following minimum description length (MDL) test statistic:
\begin{align*}
    {\rm MDL}\left( U_c,V_d \right)
    &= \log_2 \left[ \binom{n}{|U_c|} \binom{m}{|V_d|} \right] + \left( \frac{1-\mu_1^2}{2\ln2}-L_\zeta \right)|U_c|\times|V_d|,
\end{align*}
where $\mu_1$ is the mean value of edge-wise test statistics $\zeta_{ij}$
%(i.e., $\mathbf{W}$ with test statistics $\zeta_{ij}$ as entries)
for edges within ${U}_c \times {V}_d$, and $L_{\zeta}=-\int \phi(\zeta_{ij})\log_2 \left( \phi(\zeta_{ij}) \right) d\zeta_{ij}+C$ is an information entropy measure based on the standard normal distribution $\phi$ for $\zeta_{ij}$. Detailed derivations for the MDL-based test statistic and its connections to our inference goal are provided in Appendix~B.4. We formally present the cluster-wise permutation test for each observed sub-area pair $(U_c,V_d)$ in Algorithm~\ref{alg:alg3}. The number of permutations $H$ in this algorithm can be determined based on the sample size, the targeted computational expense, and the precision of the test. For example, $H=1000$. Compared to conventional multiple testing correction methods (e.g., FDR and FWER), the MDL-based cluster-wise permutation test returns suppressed false-positive findings and shows improved statistical power in real-data examples and simulations.

\begin{algorithm}[ht!]
\caption{MDL-based cluster-wise permutation test for each $(U_c,V_d)$ pair}
\label{alg:alg3}
\begin{algorithmic}[1]
\Procedure{Algorithm }{}
    \State Compute $T^0_{c,d} = {\rm MDL}\left( U_c,V_d\right)$ for each $(U_c,V_d)$ pair yielded with true covariate labels

    %\vspace{1mm}
    \For {$h=1,\ldots,H$}
           \State Permute covariate labels and obtain the new inference connectivity matrix $\mathbf{W^h}$

           \State Obtain $U^h = \bigoplus_{c=1}^C U^h_c$ and $V^h = \bigoplus_{d=1}^D V^h_d$ by substituting $\mathbf{W^h}$ in Algorithms~\ref{alg:alg1} and \ref{alg:alg2}

           \State \textbf{return} $T^h=\rm max \left({\rm MDL}\left( U^h_c,V^h_d\right)  \right)$
    \EndFor

    \State Compute $p$-value for each observed $(U_c,V_d)$ pair: $P_{c,d}=\frac{ \sum I(T^h > T^0_{c,d}) } {H}$
    %\vspace{1mm}
    \State \textbf{return} the significance of each observed $(U_c,V_d)$ pair based on $P_{c,d}$ at a predetermined $\alpha$-level

\EndProcedure
\end{algorithmic}
\end{algorithm}

%%%%%%%%%%%%%%%%%%%%%%%%%%%%%%%%%%%%%%%%%%%%%%%%%%%%%%
\section {Experiments}
In this section, we apply SCCN to two real datasets to investigate the voxel-level dysconnectivity under specific clinical settings. Dataset~1 includes 330 participants from a schizophrenia (SZ) research study using fMRI data collected in Baltimore, MD. Dataset~2 contains 3269 participants from a nicotine-addiction study using fMRI data collected from the UK Biobank database.
% For Dataset 1, SCCN identifies aberrant vFC patterns related to schizophrenic disorder within the salience network. For Dataset 2, SCCN identifies aberrant vFC patterns related to nicotine addiction between the basal ganglia, hippocampus, and insular gyrus. The results of both datasets align with previous medical findings, and they return additional more precise altered vFC information on a voxel level, which may provide complementary insights into the underlying neurophysiological mechanisms.
% We also confirm that SCCN is well scalable to large datasets of more than thousands of subjects (e.g., UK Biobank). Moreover, SCCN shows better performance on the two real datasets in controlling false positive rate and increasing true positive rate, compared to other commonly used methods.

\subsection{Schizophrenia research study}
Our primary dataset contains 330 individuals, including 148 SZ patients (M/F 84/64, age $37.5\pm 14.4$) and 182 healthy controls (M/F 80/102, age $37.0 \pm 16.1$). The participants were required for a large ongoing study of the effects of cognitive deficits in SZ. Specifically, the study probed how cognitive deficits contributed to functional disability in SZ patients and how they were related to altered functional networks that serve cognition. All subjects were assessed at local research centers in the greater Baltimore area between 2004 and 2016 using uniform recruitment criteria, and neurological and clinical assessments. Detailed information about participant demographics, the recruitment process, imaging acquisition, and fMRI preprocessing procedures can be found in Appendix~C.1.

%\tongA{Review says: "More information/discussion could be provided for the 330 schizophrenia dataset. For example, why is this dataset important and worth studying (and thus the findings here are important)?"}

%\vspace{2mm}
\subsubsection{Salience network disrupted connectivity}
\paragraph*{Clinical background}
The salience network, which is mainly composed of the bilateral insula and cingulate cortices, is related to several core SZ symptoms. A vast amount of literature in neuroimaging research suggests that the connectivity in the salience network is disturbed during information processing in SZ patients \citep{palaniyappan2012concept}. We therefore intend to focus on the bilateral insula and cingulate cortices and study the schizophrenic-altered vFC patterns between them. Specifically, we want to extract schizophrenic-impacted edges that connect voxels from spatially coherent sub-areas within the insula to those within the cingulate cortex. This data-driven extraction of sub-areas caused by vFC abnormality in SZ may provide insights for more effective clinical treatments (e.g., by transcranial magnetic stimulation or deep-brain-stimulation therapies).

We labeled the bilateral insula and cingulate cortices based on the Brainnetome Atlas \citep{fan2016human} (left insula: 1762 voxels; right insula: 1577 voxels; cingulate cortex: 5768 voxels). We first calculated the vFC matrix between the left/right insula and cingulate cortex for each subject. Each entry in the vFC matrix was a Fisher's $z$-transformed Pearson correlation coefficient. Next, we obtained the population-level statistical inference matrices $\mathbf{W}^L_{1762\times 5768}$ and $\mathbf{W}^R_{1577\times 5768}$ across all subjects; each entry in $\mathbf{W}^L$ and $\mathbf{W}^R$ is endowed with a negative $\log p$-value quantifying the evidence of vFC differences between the SZ and healthy control groups. We then applied SCCN and the MDL-based test to $\mathbf{W}^L$ and $\mathbf{W}^R$ to respectively extract aberrant sub-area pairs between the left/right insular and cingulate cortex with the spatial-contiguity constraints. Lastly, we compared the results with those obtained using popular existing methods.

\begin{figure}[htb]
\makebox[\textwidth][c]{
    \includegraphics[width=0.8\textwidth]{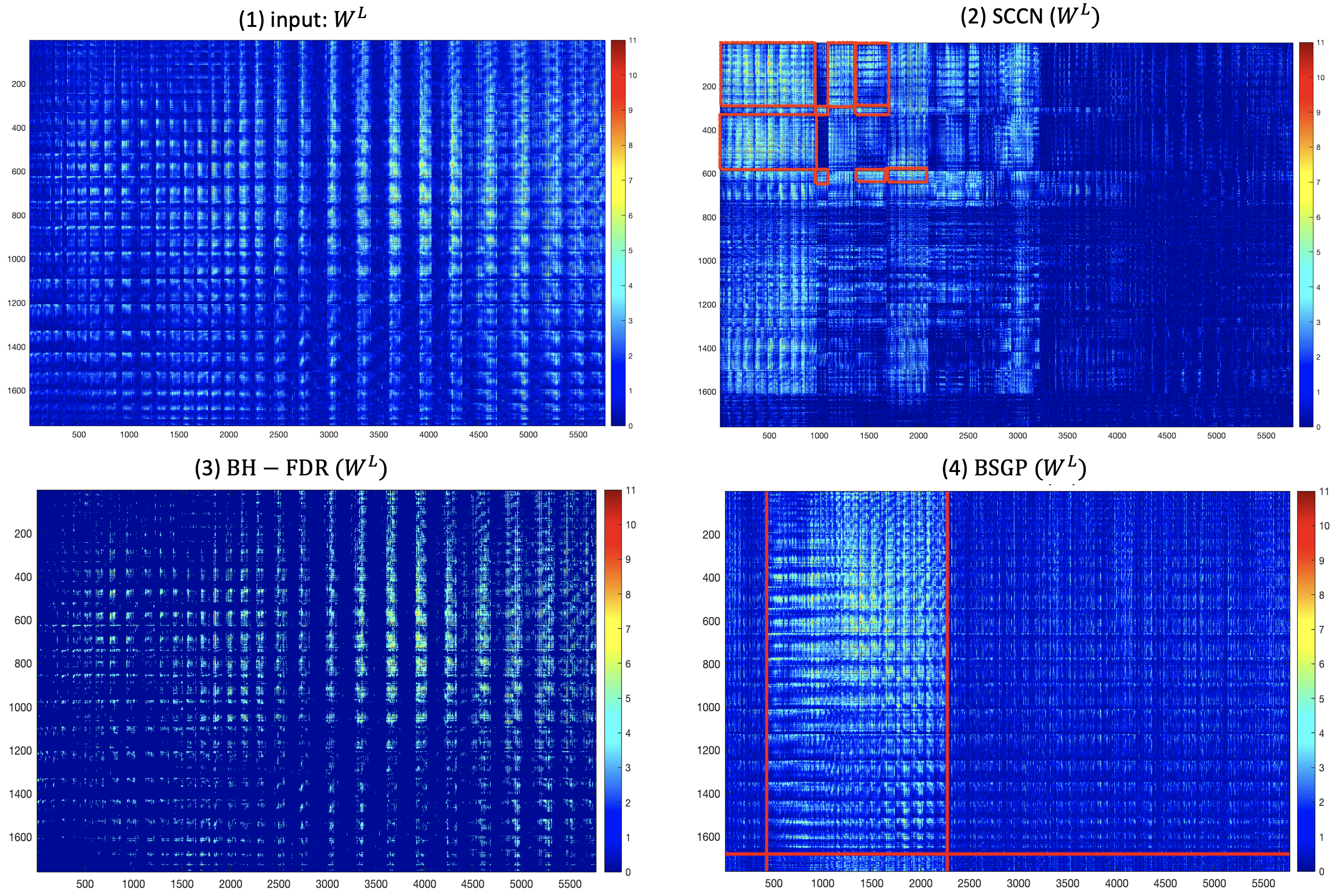}
}
% \centerline{\includegraphics[width=0.95\textwidth]{pics/WLWR1.png}}
\caption{\footnotesize \textit{(1) A heatmap of $\mathbf{W}^L$: rows and columns correspond to the voxels from the left insula and the cingulate cortex, respectively. A hotter entry indicates a more differentially expressed voxel pair between clinical groups adjusted for other covariates. (2) Results yielded by SCCN: positive sub-area pairs that pass the MDL-based permutation test are highlighted in red boxes. There are many edges with small p-values outside the red boxes (e.g., in the bottom left corner) because they are not spatially contiguous to those inside the boxes, and are automatically excluded by SCCN. (3) Results yielded by BH-FDR: with $q=0.05$, no sub-area pairs were detected. (4) Results yielded by BSGP: only one informative yet much less dense sub-area pair was detected. The detected sub-area pair was also lack of spatial contiguity and specificity.}}
\label{tab:Figure}
\end{figure}

%%%%%%%%%%%%%%%%%%%%%%%%%%%%%%%%%%%%%%%%%%%%%%%%%%%%%%
%\vspace{2mm}
\paragraph*{Network-level results}
Each element in the vFC inference matrix $\mathbf{W}^L$ is $\mathbf{W}^L_{ij}=- \log(p^L_{ij})$, where $p^L_{ij}$ is the $p$-value testing the case-control vFC difference for the $(i,j)$ pair between the left insula and cingulate cortex (\autoref{tab:Figure}(L1)). We then perform screening on $\mathbf{W}^L$ using a pre-selected threshold (e.g., $p=0.05$): $\mathbf{W}^L_{ij} = (\mathbf{W}^L)_{ij}\cdot I\left((\mathbf{W}^L)_{ij}\leq -\log(0.05)\right)$. The post-screened inference matrix $\mathbf{W}^L$ can effectively exclude most non-informative false-positive edges while maintaining a high proportion of true-positive edges \citep{fan2008sure,li2012feature}. Similar settings apply to $\mathbf{W}^R$ (\autoref{tab:Figure}(R1)). Implementing Algorithm~\ref{alg:alg2} returned a maximum-likelihood estimation (MLE) of $\hat{\lambda}_L=0.625$ for $\mathbf{W}^L$ and $\hat{\lambda}_R=0.75$ for $\mathbf{W}^R$.

\begin{figure}[htb]
\makebox[\textwidth][c]{
    \includegraphics[width=\columnwidth]{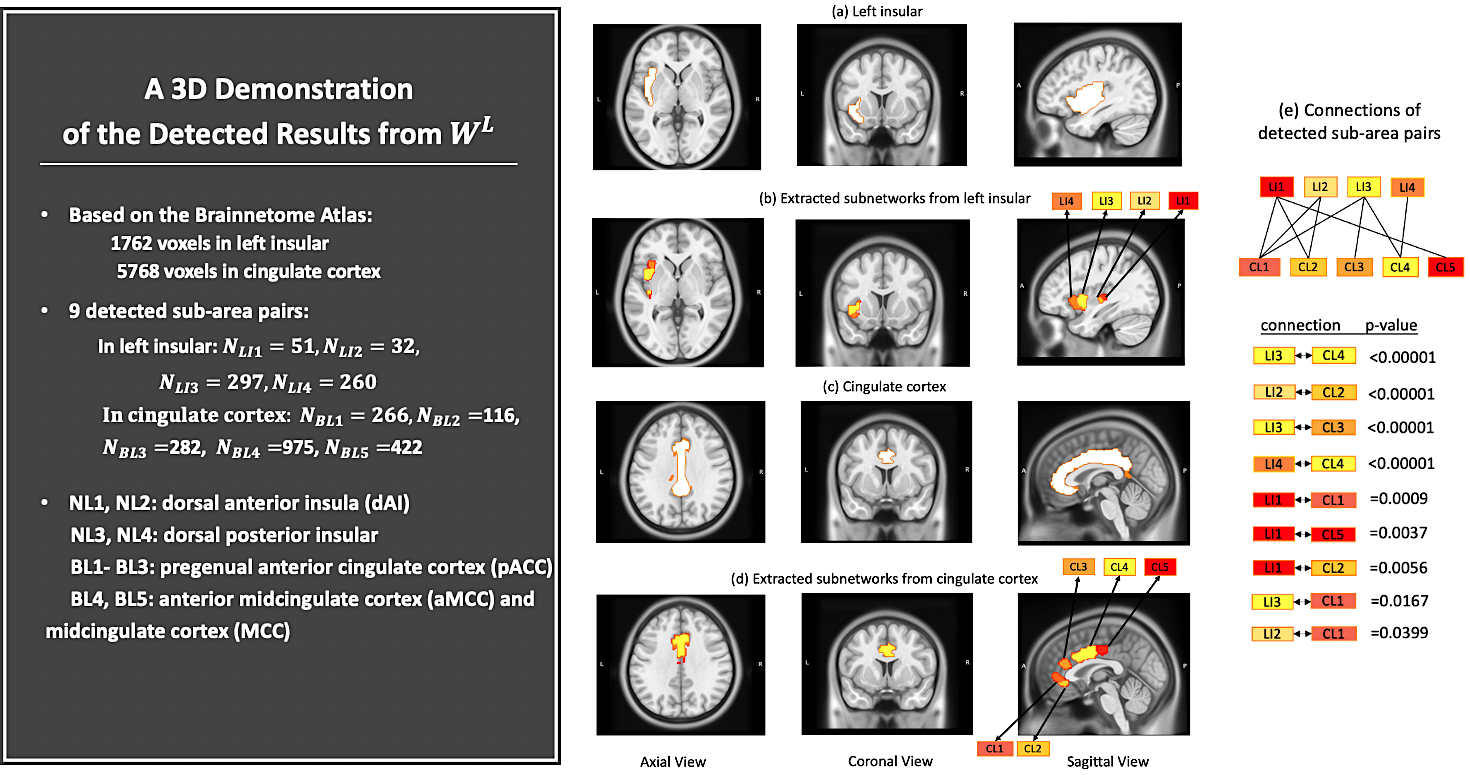}
}
% \centerline{\includegraphics[width=0.99\textwidth]{pics/left3d.png}}
\caption{\footnotesize \textit{A 3D demonstration of the detected sub-area pairs from $\mathbf{W}^L$. Let $LI_i$ be the $i$-th dysconnected sub-area detected from the left insula that is connected to the $j$-th sub-area from the cingulate cortex, $CL_{j}$. Let $N_{LI_i}$ denote the number of voxels in sub-area $LI_i$, and similarly $N_{CL_{j}}$ for $CL_{j}$. (a)(c) show the images of the original left insular and cingulate cortex; (b) shows the SZ-affected sub-areas in the left insular that are connected to those in the cingulate cortex highlighted in (d); (e) shows the architecture of interconnections between the detected sub-areas from $\mathbf{W}^L$ and the associated $p$-values from the MDL-based permutation test. A 3D demonstration of the detected results from $\mathbf{W}^R$ is provided in Appendix~C.2.
}}
\label{tab:3Ddecom}
\end{figure}

Given the estimated $\hat \lambda$, Algorithm~\ref{alg:alg1} returned the number of clusters $\hat C_L=135, \hat D_L=107$ for $\mathbf{W}^L$, and $\hat C_R=225, \hat D_R=226$ for $\mathbf{W}^R$. The MDL-based test returned nine abnormal sub-area pairs for $\mathbf{W}^L$ and ten abnormal sub-area pairs for $\mathbf{W}^R$ (marked in red in \autoref{tab:Figure}(L2) and (R2)). A 3D demonstration of the detected results from $\mathbf{W}^L$ is shown in \autoref{tab:3Ddecom} (using a significance level of $0.05$ from the MDL-based permutation test). Information regarding the precise sizes, $p$-values, and locations is also specified in \autoref{tab:3Ddecom}. All extracted sub-area pairs show well-organized topological structures. Overall, the aberrant vFC patterns from $\mathbf{W}^L$ are gathered between the dorsal insula and anterior cingulate cortex (ACC). Detailed detection results for $\mathbf{W}^R$ are provided in Appendix~C.2.

%\begin{sidewaysfigure}
%\end{sidewaysfigure}

%%%%%%%%%%%%%%%%%%%%%%%%%%%%%%%%%%%%%%%%%%%%%%%%%%%%%%
%\vspace{2mm}
\paragraph*{Biological interpretation of detected sub-areas}
%The brain region constellation of
%SCCN reveals data-driven altered vFC patterns within the salience network.
The detected sub-areas consist of several well-known brain regions that are believed to be frequently associated with SZ disorder, including, most remarkably, the anterior insula (AI) and ACC. Emotions that most strongly engage the AI, such as anger and fear, are those that SZ patients tend to have the most difficulty recognizing \citep{wylie2010role}. Furthermore, the densities of neurons, axons, and synapses are found to be abnormal in the ACCs of people with SZ \citep{arnold1996recent}. All of the aberrant edges detected showed decreased or equivalent connections in SZ patients. This aligns with medical findings that SZ is a ``dysconnectivity'' disorder with primarily reduced FC across the salience network \citep{lynall2010functional}, although medication effects cannot be completely ruled out. The imposed spatial-contiguity constraints help unfold brain sub-areas of the bilateral insula and cingulate cortices that maximally cover disease-related vFC. These novel findings improve the spatial specificity of SZ-related dysconnectivity in the well-known salience network and may lead to guidance for future treatments.

%\vspace{2mm}
\paragraph*{Comparisons with existing methods}
For comparison purposes, we performed the Benjamini--Hochberg FDR (BH-FDR) correction edge-wisely and a commonly used biclustering algorithm, bipartite spectral graph partitioning (BSGP), cluster-wisely. By first conducting an initial correlation analysis between vFC and schizophrenic status, $17.84\%$ of the edges in $\mathbf{W}^L$ were found to have $p<0.005$ significance, where $p = 0.005$ is a commonly used yet uncorrected threshold in neuroimaging studies \citep{derado2010modeling}. After applying BH-FDR correction, $9.45\%$ of the edges were found to be significant using the threshold of $q = 0.01$ (\autoref{tab:Figure}(L3)), and no community structure was revealed. For $\mathbf{W}^R$, $13.50\%$ of edges had $p$-values less than $0.005$, and only $3.61\%$ significant edges were found after BH-FDR correction with $q = 0.01$ (\autoref{tab:Figure}(R3)); again, no community structure was found in $\mathbf{W}^R$. When applying BSGP to both $\mathbf{W}^L$ and $\mathbf{W}^R$, only one abnormal sub-area pair was detected (\autoref{tab:Figure}(L4) and (R4)), with more than $36.80\%$ edges of $p>0.005$ included compared to SCCN. In comparison to the existing methods, SCCN yields much more densely schizophrenia-associated vFC contained in spatially contiguous sub-area pairs with stronger topological structures.

% \begin{sidewaysfigure}
%     \centerline{\includegraphics[scale=.30]{pics/comp_horizaontal.png}}
%     \caption{The comparisons of different methods to detect abnormal voxel-level connections}
%     \label{tab:comparisons}
% \end{sidewaysfigure}

%\vspace{2mm}
\subsubsection{Temporal-thalamic disrupted connectivity}
In contrast to the reduced salience network connections in SZ patients, many studies have shown that SZ patients have greater thalamic connectivity with multiple sensory-motor regions, including, most remarkably, the temporal gyrus \citep{ferri2018resting, cetin2014thalamus}. More specifically, thalamus to middle temporal gyrus connectivity was positively correlated with many core SZ features, such as hallucinations and delusions. We therefore aim to use SCCN to identify some novel findings between the middle temporal gyrus on the right hemisphere and the bilateral thalamus in SZ patients. Based on the Brainnetome Atlas, there are 3566 voxels in the right middle temporal gyrus (labeled $82$, $84$, $86$, and $88$) and 3275 voxels in the bilateral thalamus (labeled $231$--$246$). We computed the vFC connectivity inference matrices $\mathbf{W}^{\mathrm{(Tem_{right}, Tha_{left})}}_{3566\times 1727}$ and $\mathbf{W}^{\mathrm{(Tem_{right}, Tha_{right})}}_{3566\times 1548}$ between clinical groups and then implemented SCCN. Due to limited space here, we provide the results for the selections of all parameters and densely altered sub-area pairs in Appendix~C.3.

%%%%%%%%%%%%%%%%%%%%%%%%%%%%%%%%%%%%%%%%%%%%%%%%%%%%%%
\subsection{Nicotine-addiction research study}
Our primary dataset contains 3269 individuals from the UK Biobank database, including 1353 constant current smokers (M/F: 2653/616, age: $48.6\pm 15.3$) and 1916 previous light smokers (M/F: 1187/729, age: $32.9 \pm 18.1$). Specifically, we define current smokers as participants who currently smoke more than ten cigarettes per day (i.e., cases who are addicted to nicotine).
%\footnote{ACE touchscreen question "About how many cigarettes do you smoke on average each day? "}.
We define previous light smokers as those who indicated that they had only tried a few cigarettes in the past but are not currently addicted to nicotine products (i.e., controls).
%\footnote{ACE touchscreen question "In the past, how often have you smoked tobacco?"}.
% \tblue{We exclude those who used to smoke on most or all days but quit because this may be caused by some irresistible reasons such as illness or pregnancy. We also exclude those who had never smoked in the past since those participants were never exposed to nicotine due to religious, family reasons, etc., and therefore never had a chance to become addicted to nicotine.}
By investigating different neural-connectivity patterns across the two groups, we may obtain more information on the inherent neurological mechanism of nicotine dependence and thereby help smokers resist nicotine cravings.
% }

%\vspace{2mm}
\subsubsection*{Clinical background}
Abundant literature shows that the basal ganglia (BG), hippocampus (Hippo), and insular gyrus (Ins) play important roles in nicotine addiction \citep{ersche2011abnormal, gaznick2014basal,mcclernon2016hippocampal}. We therefore intend to look into the disrupted connectivity patterns between all possible pairs (12 in total) formed by these three bilateral ROIs: (BG, Hippo), (BG, Ins), and (Hippo, Ins). To keep the presentation concise, we will present the results of the (left BG, left Ins) pair here and the remaining 11 cases in Appendix~D. Again, we labeled the left BG and left Ins using the Brainnetome Atlas (left BG: 2345 voxels; left Ins: 1762 voxels). We followed the same computational procedures as in Dataset~1 and obtained the edge-wise connectivity inference matrices $\mathbf{W}^{\mathrm{(BG_{left}, Ins_{left})}}_{2345\times 1762}$ across clinical groups. We then applied SCCN and the MDL-based test to $\mathbf{W}^{\mathrm{(BG_{left}, Ins_{left})}}$ to extract abnormal sub-area pairs with the spatial-contiguity constraints. Lastly, we compared the results with those obtained using two common existing methods.

%\vspace{2mm}
\subsubsection*{Network-level results}
\begin{figure}[htbp]
\captionsetup[subfigure]{labelformat=empty}
\subfloat{%
  \includegraphics[clip,width=0.55\columnwidth]{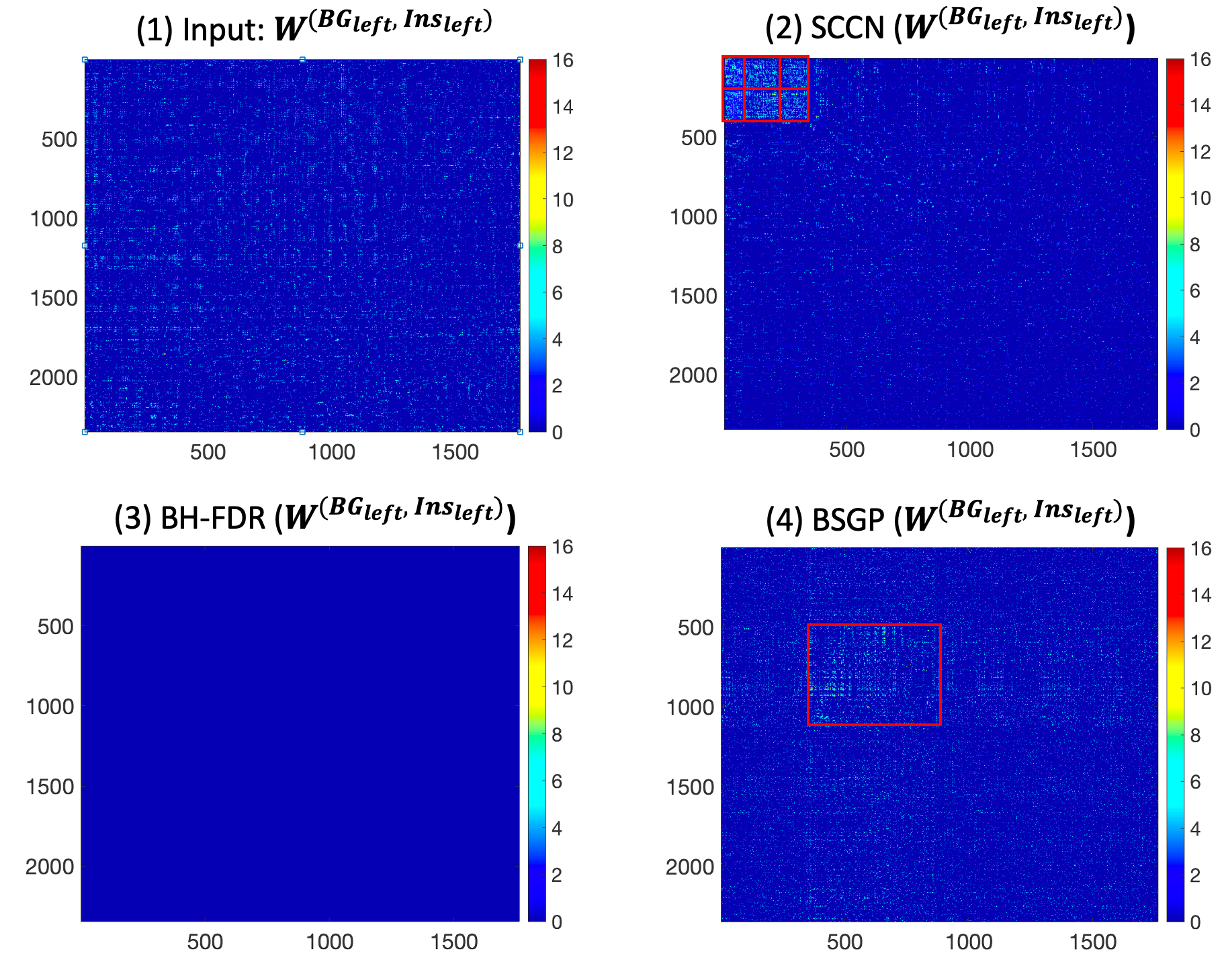}%
}

\subfloat{%
  \includegraphics[clip,width=\columnwidth]{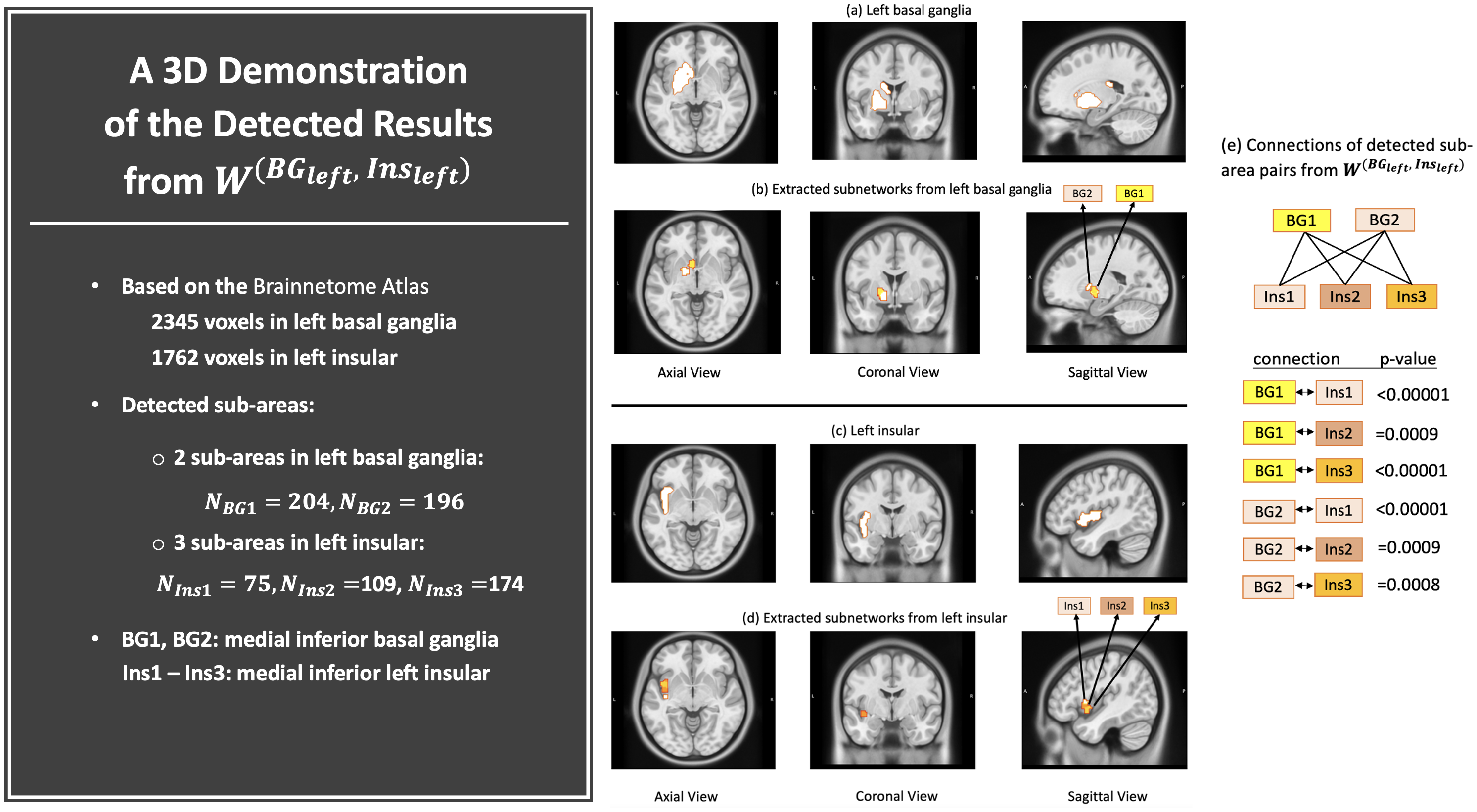}%
}

\caption{\footnotesize \textit{(1) A heatmap of $\mathbf{W}^{\mathrm{(BG_{left}, Ins_{left})}}$: rows and columns correspond to voxels from the left basal ganglia and the left insular, respectively. (2) Results yielded by SCCN: altered sub-area areas that pass the MDL-based permutation test are highlighted in red boxes. (3) Results yielded by BH-FDR: The hypothesis testing error measure was set to be $q=0.05$ as a cut-off. No sub-area pairs were detected. (4) Results yielded by BSGP: only one positive yet much less dense sub-area pair was detected. The detected sub-area pair also lack spatial contiguity and specificity. (a)-(d) shows the 3D demonstration of the 6 detected altered sub-areas from $\mathbf{W}^{\mathrm{(BG_{left}, Ins_{left})}}$. (a)-(e) show a 3D demonstration of the detected results from $\mathbf{W}^{\mathrm{(BG_{left}, Ins_{left})}}$. Based on the p-values from the MDL-based permutation test shown in (e), most positive sub-area pairs are located in the medial inferior part of left basal ganglia and left insular.}}
\label{tab:ukb}
\end{figure}

Each entry in the inference matrix $\mathbf{W}^{\mathrm{(BG_{left}, Ins_{left})}}$ is endowed with a $-\log p$ value testing the vFC difference between clinical groups. (\autoref{tab:ukb}(1)). Implementing Algorithm~\ref{alg:alg2} returned the MLE $\hat{\lambda}=0.75$. Given the estimated $\hat \lambda$, Algorithm~\ref{alg:alg1} returned the number of clusters $\hat C=306, \hat D=210$ for $\mathbf{W}^{\mathrm{(BG_{left}, Ins_{left})}}$. The MDL-based test returned six abnormal sub-area pairs, which are marked in red in \autoref{tab:ukb}(2). A 3D demonstration of the detected sub-area pairs from $\mathbf{W}^{\mathrm{(BG_{left}, Ins_{left})}}$ is shown in \autoref{tab:ukb}(a)--(e) (with a significance level of $0.05$ selected for the MLD-based permutation test). All extracted sub-area pairs show well-organized topological structures. The majority of aberrant vFC patterns from $\mathbf{W}^{\mathrm{(BG_{left}, Ins_{left})}}$ are gathered between the medial inferior part of the left basal ganglia and the left insula.

% \begin{figure}[htb]
% \makebox[\textwidth][c]{
%     \includegraphics[width=0.75\textwidth]{pics/ukb_w.png}
% }
% % \centerline{\includegraphics[width=0.95\textwidth]{pics/WLWR1.png}}
% \caption{}}
% \end{figure}

\subsubsection*{Biological interpretation of detected sub-areas}
The detected sub-areas consist of several locations that are believed to be frequently associated with nicotine addiction, including the medial inferior part of the basal ganglia and the posterior insula. We also observed decreased connectivity within these regions in current smokers, which aligns with the previous medical discovery that decreased resting-state functional connectivity is correlated with increased nicotine-addiction severity \citep{fedota2015resting, sutherland2018functional}. The incorporated spatial-contiguity constraints help unfold the sub-areas within the BG, Hippo, and Ins, which maximally cover addiction-related vFC. These novel findings improve the spatial specificity of addiction-related locations in the three brain regions and may lead to future guidance for resisting the urge to use nicotine products.

%\vspace{2mm}
\subsubsection*{Comparisons with existing methods}
For comparison purposes, we again performed the BH-FDR correction edge-wisely and BSGP cluster-wisely on $\mathbf{W}^{\mathrm{(BG_{left}, Ins_{left})}}$. By first conducting an initial edge-wise significance test across the current and previously light smoker groups, only $7.29\%$ of the edges were found to be significant ($p<0.005$). However, no edges showed significance after applying BH-FDR correction with $q = 0.01$ (\autoref{tab:ukb}(3)). When applying BSGP to $\mathbf{W}^{\mathrm{(BG_{left}, Ins_{left})}}$, only one abnormal sub-area pair was detected (\autoref{tab:ukb}(4)), with $49.5\%$ edges of $p>0.005$ included in the detected pair, compared to $3.12\%$ yielded by SCCN. In comparison to the two existing methods, SCCN yields much more densely altered vFC contained in spatially contiguous sub-area pairs with strong topological structures.

%%%%%%%%%%%%%%%%%%%%%%%%%%%%%%%%%%%%%%%%%%%%%%%%%%%%%%
\section{Simulations}
% \revision{In the simulation study, we consider two following analyses: 1. \textbf{Primary analysis}: when connections between two brain regions are believed to be related to certain brain diseases, we want to evaluate how well SCCN can detect aberrant patterns inside; 2. \textbf{Negative control analysis}: if some brain regions are known to be unrelated to certain brain conditions, we want to verify that applying SCCN will not yield false-positive sub-area pairs.}
In the simulation study, we probed whether SCCN can extract densely altered sub-area pairs with better performance compared to common existing methods. Specifically, we evaluated the performance from two perspectives. (i)~Multivariate edge-level inference: whether extracted voxel pairs have a high true-positive rate (TPR) and low false-positive rate (FPR); (ii)~network-level inference: whether the extracted sub-areas contain maximal true-positive voxels, compared to other unextracted sub-areas.

% \tongA{For the 1000 simulations with each different ($\sigma, S$), all of them yield $\mathbb{P}(\hat{U}_c \subset U_c ~\mathrm{or}~ U_c \subset \hat{U}_c) < 2\times 10^{-18}$,and $\mathbb{P}(\hat{V}_d \subseteq V_d ~\mathrm{or}~ V_d \subset \hat{V}_d) < 2\times 10^{-17}$, where $c=1,2, ~d=1,2,3,$ and $\hat{U}_c, \hat{V}_d$ are the extracted sub-areas. The network-level findings are verified to be reproducible, and we therefore shall show results from a random 1/1000 simulation with different settings ($\sigma, S$) as below. }
\begin{figure}[htbp]
\makebox[\textwidth][c]{
    \includegraphics[width=1.\textwidth]{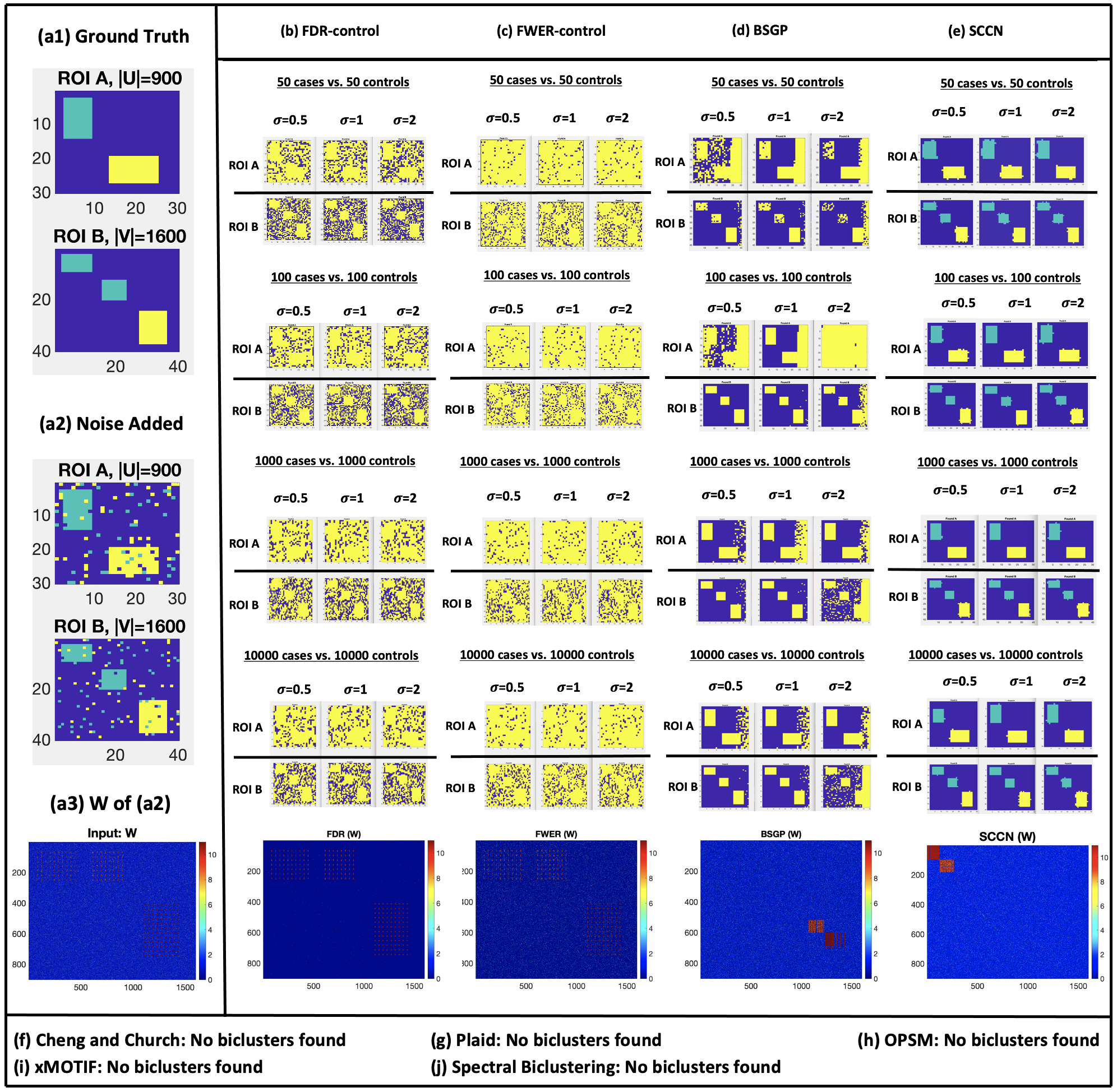}
}
% \centerline{\includegraphics[width=0.99\textwidth]{pics/simComp2.png}}
\caption{\footnotesize \textit{Visualization of simulation results. (a1) shows the ground truth locations of the disease-related sub-area pairs $(U_1, V_1)$, $(U_1, V_2)$, and $(U_2, V_3)$ with spatial contiguity (only regions with same color exhibit dysconnectivity from A to B). (a2) includes additional noise and positive but isolated voxel pairs based on (a1) to mimic the real vFC patterns in the brain connectome. (a3) shows the connectivity inference matrix $\mathbf{W}$ obtained based on (a2). (b)-(e) show the detected disease-related voxel pairs (again only regions with the same color form a pair) under different variances $\sigma$ and sample sizes $S$. The last row shows the isomorphic graphs of (a3)with the extracted sub-area pairs pushed to the top. We highlight the voxels from the supra-threshold voxel pairs that were yielded by the FDR-control and FWER-control, and voxels in sub-area pairs that were extracted by BSGP and SCCN. Multiple testing with FDR-control and FWER-control tend to extract an excess of voxels with high false-positive error rates. BSGP better controls the error rates, but it extracts voxel pairs without differentiating the correct area-wise connections, i.e., $(U_1, V_1)$, $(U_1, V_2)$, and $(U_2, V_3)$. In contrast, SCCN can simultaneously recover the spatially-contiguous sub-areas, respectively, in A and B, and reveal the correct disease-related vFC patterns. (f)-(j) show that no single differentially expressed sub-area pair was extracted by the biclustering algorithms listed.}}
\label{tab:3Dsimu}
\end{figure}

\subsection{Primary analysis}
We first generated a bipartite graph $G=\{U, V\}$ to represent the brain connectome between two brain regions~A and B for $S$ subjects (\autoref{tab:3Dsimu}(a1)), where $U$ corresponds to the voxel set in Region~A, and $V$ corresponds to that in Region~B. We assume all $S$ subjects share common node sets after spatial normalization and registration, i.e., $(U^s,V^s)\equiv (U,V), \forall s\in [S]$. Next, we simulated covariates of interest $\{\mathbf{X}^1,\ldots,\mathbf{X}^S\}$ that contain clinical information of all $S$ subjects. Lastly, we simulated the Fisher's $z$-transformation connectivity matrices $\{\mathbf{Z}^1, \ldots, \mathbf{Z}^S\}$ between regions~A and B for all subjects, where $\mathbf{Z}^s \in \mathbb{R}^{n \times m}, n=|U|, m=|V|$. Specifically, each element $z^s_{ij}$ in $\mathbf{Z}^s$ was set to follow $\mathcal{N} (h(z^s_{ij}),\sigma^2)$, where $h(z^s_{ij})=\mathbf{X}^s \mathbf{\beta}_{ij}$ is location-specific within regions~A and B.

In the following, we show the numerical settings under the above simulation framework:
\begin{enumerate}
\item For the two pre-defined brain regions of interest, we simulated $|U|=900$ voxels in Region~A and $|V|=1600$ voxels in Region~B. Within $|U|$ and $|V|$, we also randomly simulated three disease-related sub-area pairs $(U_1, V_1)$, $(U_1, V_2)$, and $(U_2, V_3)$. Not every possible pair $\{(U_c, V_d), c=[2], d=[3]\}$ was associated with the disease; only regions with the same color exhibited dysconnectivity from A to B (see \autoref{tab:3Dsimu}(a1)). The sizes of these sub-area pairs were $|U_1||V_1| =84\times 70=5880$, $|U_1||V_2|=84\times 64=5376$, and $|U_2||V_3|=96\times 117 =11\,232$. In addition, we included spatially isolated abnormal voxels as well as noise within regions~A and B to mimic more realistic neural connectivity (\autoref{tab:3Dsimu}(a2)).

%\vspace{2mm}

\item For the Fisher's $z$-transformation connectivity matrices $\{\mathbf{Z}^s, s\in S\}$, we set $h(z^s_{ij})=\beta_0+\beta_{{ij},1}x^s_1+\beta_{{ij},2}x^s_2+\beta_{{ij},3}x^s_3$, where $x^s_1$ and $x^s_2$ store the age and sex information for subject $s$, and $x^s_3$ represents their clinical status ($x^s_3$=1 if patient $s$ has a mental disorder, and $0$ for a healthy control.). In addition, while $\beta_{{ij},1}$ and $\beta_{{ij},2}$ are typically not spatially variant, $\beta_{{ij},3}$ is considered brain-region specific:
\begin{equation*}
    \beta_{ij,3}=\left\{
        \begin{array}{ll}
        0.9, & \text{if}\quad (i,j) \in (U_1,V_1) ~\bigcup~ (U_1,V_2), \\
        0.13, & \text{if}\quad (i,j) \in (U_2, V_3), \\
        0, & \text{if}\quad (i,j) \in U/\{ (U_1,V_1) ~\bigcup~ (U_1,V_2) ~\bigcup~ (U_2, V_3) \}.
        \end{array}
        \right.
\end{equation*}

%\vspace{2mm}
\item To control standardized effect sizes, we set $\sigma^2=0.5, 1.0, 2.0$ in $\mathbf{Z}^s \allowbreak \sim \mathcal{N} (h(z^s_{ij}),\allowbreak \sigma^2)$. Additionally, four sample sizes, $S=100$, $200$, $2000$, and $20,000$, were used, each with balanced healthy controls and patients. All settings with different ($\sigma, S$) were simulated for 1000 times to assess the variability of the TPR and FPR.
\end{enumerate}

We implemented Algorithm \ref{alg:alg1} and \ref{alg:alg2} of SCCN to identify sub-area pairs from each simulated dataset, and we then applied Algorithm~\ref{alg:alg3} to conduct cluster-wise inference on the sub-area pairs detected. To assess the performance of the multivariate edge-wise inference, we considered two conventional multiple-testing controls (FDR and FWER). Specifically, we used the voxel-wise permutation test (with 1000 permutations) to control the FWER and the Benjamini--Hochberg procedure (with $q=0.05$ as a cut-off) to control the FDR \citep{benjamini1995controlling}. To assess the accuracy of the cluster-wise performance, our goal was to compare true disease-related subgraphs $\{(U_c, V_d)\}$ with the estimated subgraphs $\{(\hat{U}_c, \hat{V}_d)\}$ produced by five commonly used biclustering algorithms (i.e., Cheng and Church, Plaid, OPSM, xMOTIF, and Spectral Biclustering \citep{verma2013ranking}).
%to benchmark the performance of SCCN.

The edge-wise inference results are presented in Table~1, and a graph illustration of the results is shown in \autoref{tab:3Dsimu}. For the edge-wise inference performance with all different $(\sigma, S)$, SCCN outperforms the two traditional multiple testing correction methods (i.e., FDR and FWER control) in terms of TPR, while its ability to control the FPR falls in between the two. SCCN's relatively inferior performance in controlling the FPR (compared to sensitivity) can sometimes be impacted by the following disadvantage: in traditional multiple testing methods with universal thresholds, one false-positive finding corresponds to exactly one false-positive edge. However, SCCN detects altered edges by partitioning voxels within each ROI; therefore, one false-positive finding by SCCN corresponds to one false-positive voxel, say $v_i \in U_c$, which will lead to $n$ false positive edges when $V_d$ ($|V_d|=n$) is found to connect to $|U_c|$. The greater the size of $V_d$, the more false-positive edges will be yielded. Nonetheless, even with such a heavy penalty for detecting one false-positive voxel, SCCN still controls the FPR and shows better performance when jointly considering the TPR and FPR.
% Also, SCCN shows superior performance in extracting sub-networks with community structures, of which the two traditional multiple testing methods are incapable.
More importantly, false-positive edges discovered by the traditional FDR and FWER correction approaches almost cover all within-ROI voxels, which leads to a substantial loss of spatial specificity when identifying covariate-related vFC patterns.

Regarding the network-level inference performance, all common biclustering methods failed to detect any positive biclusters (differentially expressed sub-area pairs) except for BSGP. However, BSGP nonetheless failed to ensure spatial contiguity, and the precise connection between the extracted sub-areas was not correctly revealed. That is, unlike the results yielded SCCN (\autoref{tab:3Dsimu}(e)), BSGP (\autoref{tab:3Dsimu}(d)) could not effectively differentiate between yellow and blue clusters. In comparison, SCCN shows outstanding network-level performance for detecting community structures and incorporating spatial contiguity.

\captionsetup{labelformat=empty}
\begin{figure}[htbp]
\makebox[\textwidth][c]{
    \includegraphics[width=1\textwidth]{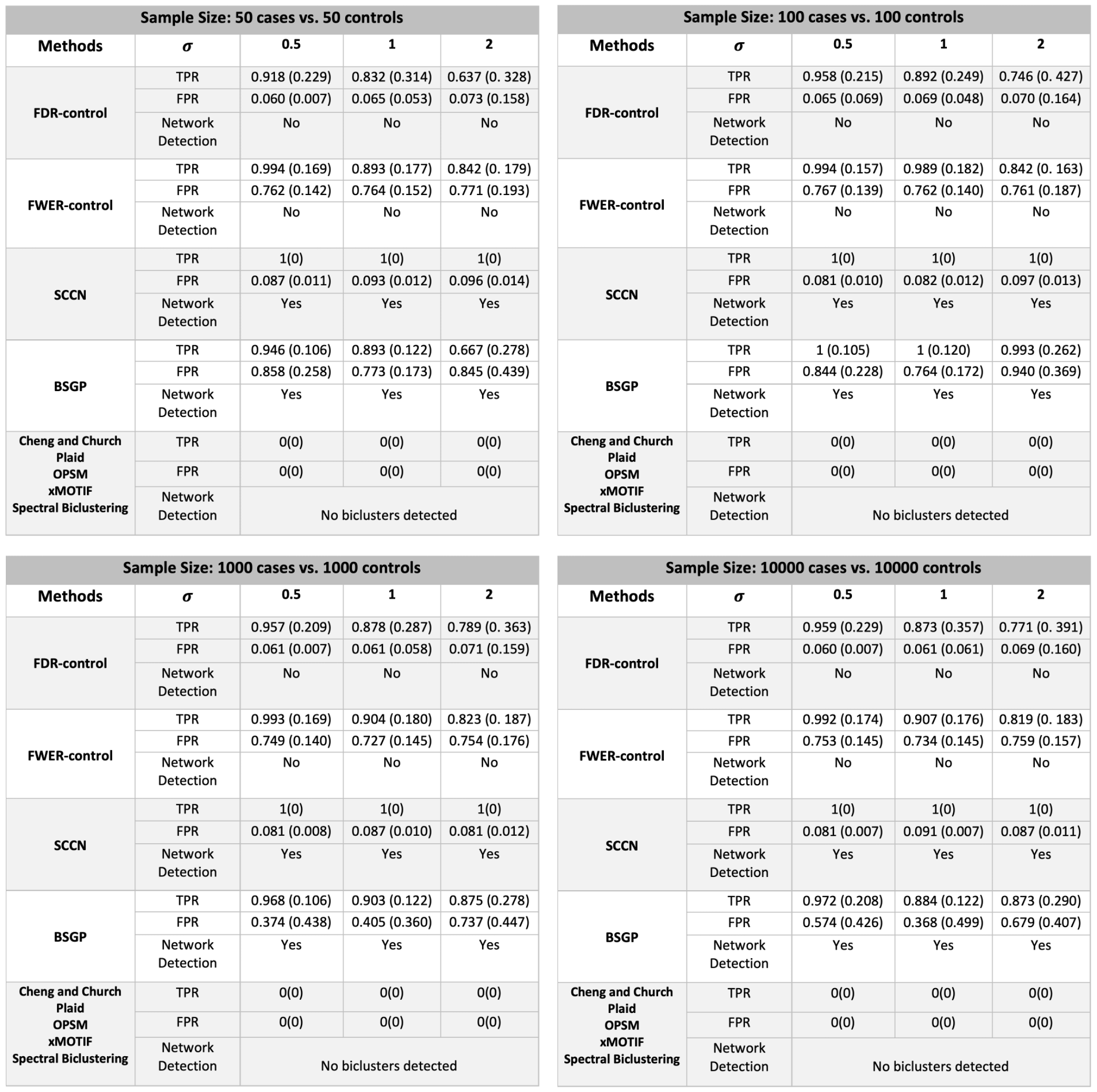}
}
% \centerline{\includegraphics[width=0.99\textwidth]{pics/simComp2.png}}

\caption{\footnotesize \textit{Table 1: Simulation results. The four sub-tables show the inference results given different sample sizes and variances, where TPR and FPR correspond to the edge-wise true positive rate and false positive rate. Network detection results indicate whether the algorithm can successfully extract the correct connection patterns between disease-related sub-area pairs.}}
\label{tab:simu_table}
\end{figure}

\subsection{Negative control analysis}
We further performed a negative control analysis to evaluate the FPR of our method. We consider a scenario in which the connections between a pre-selected ROI pair are unrelated to a clinical condition of interest. We generated $|U|=900$ and $|V|=1600$ voxels in regions~A and B. We distinguished the patient and control groups as $1$ and $0$, but since there were no abnormal sub-area pairs $\{(U_c, V_d)\}$ across groups, we simply set the connectivity matrices $\mathbf{Z}^s \sim \mathcal{N} (0,\sigma^2)$ over the entire regions for all $S$ subjects. Based on $\mathbf{Z}^s$, we obtained the inference matrix $\mathbf{W^0}$ across clinical groups. Since the network detection was validated to be scalable to different sample sizes and sample variances, we evaluated the configuration $(S=1000, \sigma=1)$ as a proof of concept. Finally, we implemented SCCN on $\mathbf{W^0}$. 
Since the false positive voxel pairs tended to be distributed randomly, no sub-area pairs were significant.
Therefore, the sub-area-level false positive findings were 0. 
The edge-wise FPR (supra-threshold voxel-pairs) among 1000 iterations was $6.82\times10^{-5} (\mathrm{std.}~ 1.29\times10^{-5})$, which with consistent with the pre-determined alpha level ($\mathbb{E}(p)=0.00005$). We have provided a graph visualization of these results in Appendix E. 
%\vspace{2mm}

In summary, we have shown that the sub-area detection is not affected by different values of variance $\sigma^2$, sample size $S$, or other sources of noise. SCCN also yields vFC patterns with high sensitivity and low FPRs. The spatial-contiguity constraints allow positive edges to borrow strengths from each other within a data-driven sub-area; sensitivity is thus notably increased. Data-driven sub-areas with these constraints can also exclude false-positive edges that bridge voxels that are randomly scattered in ROIs. False-positive findings are therefore largely suppressed. In addition, the jointly improved sensitivity (and thus statistical power) and control of the FPR yield almost identical voxel sets across all simulated datasets. Replicability is hence remarkably improved.

\section{Discussion}
Psychiatric and neurological disorders are often associated with a disrupted brain connectome. To improve the spatial specificity and sensitivity for detecting a disease-impacted brain connectome, in this work, we focused on voxel-level connectivity network analysis. We developed statistical models focusing on extracting abnormal voxel pairs from a region pair of interest, which can be further extended to whole-brain connectome analysis. We have attempted to simultaneously address the challenges of a controlled FPR for multiple voxel-pair testing and the spatial-contiguity constraints for vFC analysis. In fact, it is also possible that disease-related voxel-level connectivity occurs within a region. We can apply a similar approach to a pre-obtained within-region adjacency matrix $\mathbf{W}_{n\times n}$ by integrating a corresponding spatial-contiguity infrastructure graph, say ${\cal S}_A$. Following this, SCCN becomes applicable to intra-region voxel-level connectivity analysis. We provide detailed procedures for this in Appendix~A.3.
In addition, the brain parcellation to extract sub-areas is usually based on commonly used brain atlases (e.g., Brodmann's map or the International Consortium for Brain Mapping), and these were built on comprehensively studied cortical anatomy, such as complex gyro-sulcal folding patterns. Different regions blocked by gyri and sulci tend to show differential neurobiological structures and functions, and these atlases can thus serve as a good foundation to investigate sub-area community structures. However, to further overcome the limitation of using existing brain parcellations, one can consider combining any extracted spatially adjacent sub-areas from a pair of spatially adjacent regions if the combination is statistically coherent and biologically meaningful.

The centerpiece of our proposed method is the identification of sub-area pairs containing an unusually high density of phenotype-related voxel pairs. By leveraging this high density, we can effectively control the FPR by excluding isolated false-positive edges, and we thus greatly reduce the number of false-positive nodes. We have therefore improved the spatial specificity of extracted disease-related patterns at a voxel level. Herein, we have proposed a new non-parametric objective function to achieve this goal, and this has been implemented with efficient algorithms. We also developed inference methods to assess the statistical significance of each sub-area pair extracted.

The biological findings from our data example are novel; SCCN revealed vFC connectome patterns for schizophrenia within the well-known salience network. We discovered that the malfunction of salience network connectivity is mainly driven by disrupted connections between the dorsal insula and anterior cingulate cortex instead of the omnibus region-level findings. We further validated our findings through extensive simulations and showed that our methods could improve sensitivity with a controlled FPR while retaining spatial contiguity.

In summary, SCCN provides a new toolkit for vFC analysis with improved spatial resolution and specificity while preserving a well-controlled false-positive error rate. Therefore, the findings from SCCN can be translated into more effective potential treatments for brain disorders. Since the input data of SCCN is voxel-pair-level inference results, it is applicable to all connectivity measures and data modalities where valid statistical inference can be performed (e.g., white-matter tractography). SCCN may also provide a promising strategy for whole-brain connectome voxel-pair network analysis. All sample code can be found at \href{https://github.com/TongLu-bit/DecodingNetwork}{https://github.com/TongLu-bit/DecodingNetwork}.

%\tcyan{ We are optimistic that SCCN opens a new avenue for the whole-brain voxel-pair connectivity network analysis, which could bring promising guidance for medical assessments. Besides charaterizing neuron connectivity, SCCN can also be applied to other brain connectome study such as identifying DTI modality structures. Moreover, while SCCN was originally devised for use in brain connectomics, the method is suited to quasi-biclique mining in many other networks such as social, economic, and telecommunication networks. }

%\vspace{8mm}

% \textbf{Acknowledgments.}
% The authors would like to thank the anonymous referees, the Associate Editor, and the Editor for their constructive comments that improved the quality of this paper.
% \begin{funding}
% Funding for the project was provided by the National Institute on Drug Abuse of the National Institutes of Health under Award Number 1DP1DA048968-01.
% \end{funding}

%\vspace{5mm}
%\clearpage
\nocite{*} % to test all bib entrys
\bibliographystyle{apalike}

\bibliography{Reference}
\end{document}